\documentclass[twocolumn,3p,times,procedia]{elsarticle}

\usepackage{ecrc}
\usepackage{graphicx}
\usepackage{balance}
\usepackage{array,threeparttable}
\usepackage{url}
\usepackage{booktabs}
\usepackage{multirow}
\usepackage{color}

\volume{00}

\firstpage{1}

\runauth{}

\jid{procs}

\CopyrightLine{2022}{Published by Elsevier Ltd.}

\usepackage{amssymb}
\usepackage{amsmath}
\usepackage{multicol}
\usepackage{algorithm}
\usepackage{algpseudocode}
\usepackage{subfigure}

\usepackage[figuresright]{rotating}
\usepackage{bm}
\usepackage{caption}
\usepackage{titlesec}
\titleformat*{\section}{\small \bf}
\titleformat*{\subsection}{\small \em}
\titleformat*{\subsubsection}{\small \em}

\usepackage{fancyhdr}
\fancyheadoffset[RO,EL]{0pt}
\usepackage{geometry}
\begin{document}\small
\begin{frontmatter}




\dochead{}
\title{
\begin{flushleft}
{\LARGE Interest-Aware Joint Caching, Computing, and Communication Optimization for Mobile VR Delivery in MEC Networks}
\end{flushleft}
}
 %

\author[]{ \leftline {Baojie Fu$^{a,b,c}$, Tong Tang$^{a,b,c}$, Dapeng Wu$^{*,a,b,c}$, Ruyan Wang$^{a,b,c}$}}

\address{ \leftline {$^a$School of Communications and Information Engineering, Chongqing University of Posts and Telecommunications, Chongqing 400065, China}

  \leftline{$^b$Advanced Network and Intelligent Connection Technology Key Laboratory of Chongqing Education Commission of China, Chongqing 400065, China}

  \leftline {$^c$Chongqing Key Laboratory of Ubiquitous Sensing and Networking, Chongqing 400065, China}

}

\cortext[]{Corresponding author.}

\fntext[]{E-mail address: d210101003@stu.cqupt.edu.cn (B. Fu), tangtong@cqupt.edu.cn (T. Tang),
wudp@cqupt.edu.cn (D. Wu), wangry@cqupt.edu.cn (R. Wang).}

\begin{abstract}

  In the upcoming B5G/6G era, virtual reality (VR) over wireless has become a typical application, which is an inevitable trend in the development of video.
  However, in immersive and interactive VR experiences, VR services typically exhibit high delay, while simultaneously posing challenges for the energy consumption of local devices.
  To address these issues, this paper aims to improve the performance of the VR service in the edge-terminal cooperative system.
  Specifically, we formulate a problem of joint caching, computing, and communication VR service policy, by optimizing the weighted sum of overall VR delivery delay and energy consumption of local devices.
  For the purpose of designing the optimal VR service policy, the optimization problem is decoupled into three independent subproblems to be solved separately.
  To enhance the caching efficiency within the network, a bidirectional encoder representations from transformers (Bert)-based user interest analysis method is first proposed to characterize the content requesting behavior accurately.
  On the basis of this, a service cost minimum-maximization problem is formulated with consideration of performance fairness among users.
  Thereafter, the joint caching and computing scheme is derived for each user with given allocation of communication resources while a bisection-based communication scheme is acquired with the given information on joint caching and computing policy.
  With alternative optimization, an optimal policy for joint caching, computing and communication based on user interest can be finally obtained.
  Simulation results are presented to demonstrate the superiority of the proposed user interest-aware caching scheme and the effective of the joint caching, computing and communication optimization policy with consideration of user fairness.
  Our code is available at \textcolor{magenta}{https://github.com/mrfuqaq1108/Interest-Aware-Joint-3C-Optimization}.

\end{abstract}


\end{frontmatter}


\section{Introduction}

Growth of mobile data is fuelled by continuous development of computing and communication technologies.
According to Ericsson's latest report, the total global mobile data traffic for each month is expected to reach 325 EB by the end of 2028, 3.6 times than 2022 counterpart \cite{ericsson}.
Driven by service uptake from a handful of global streaming providers such as Youtube, Instagram, TikTok and Netflix, video traffic is gradually increasing and accounts for the majority of global mobile data traffic, which is expected to account for 80 percent of the latter one by the end of 2028.
In recent years, VR video is becoming more and more popular compared to those traditional videos due to its realism, immersive and strong interactivity.
Audiences can experience scenes from VR videos in an immersive way through head-mounted display devices such as Meta Quest 2 \cite{meta}, Sony PlayStation VR2 \cite{playstation}, Valve Index VR Kit \cite{valve}, HTC Vive Pro 2 \cite{vive}, Meta Quest Pro \cite{metapro}, HP Reverb G2 \cite{hp} and so on.
Meanwhile, VR videos can be applied in many areas \cite{huawei}, including education \cite{rojas2023systematic}, healthcare \cite{liu2022virtual}, shopping \cite{xi2021shopping}, tourism \cite{li2021study} and esports arena \cite{turkay2021virtual}.

Compared to traditional video services, in order to ensure the interactivity and immersion, VR service requires the processing of larger amounts of data and places higher demands on data transmission and processing in the communication networks, which may lead to excessive delay and energy consumption in the network, respectively.
However, VR services are generally delay sensitive.
Specifically, the motion-to-photon (MTP) delay of a VR service elapses from the sensor detects a hand or head movement to the new image is rendered and displayed to the screen \cite{10175638}.
If the MTP delay of VR service cannot be met, the user may suffer from nausea, dizziness and motion sickness, which will seriously affect the user's experience.

It is impractical for raw VR content to be processed by local device or cloud server alone.
VR services performed by local device alone may result in significant computing delay and energy consumption due to its limited computing capability and battery, respectively, while performed by cloud server alone may result in large communication delay due to its long communication links.
Thus, mobile edge computing (MEC) has been proposed as an effective network architecture concept for further delay reduction and energy saving in VR service via enabling computing capabilities at the edge of wireless networks \cite{du2017computation}.
Offloading and computing the raw VR content at the edge server can greatly decrease the energy consumption of local devices, accelerate the computation process compared to local computing and shorten the communication link to reduce the communication delay compared to cloud computing.

In addition, caching raw VR contents on the local device in advance can also effectively reduce the communication delay due to the fact that there is no longer a need to transfer the raw VR content from cloud server to local device.
Among all the raw contents, with limited cache capacity of local device, caching the contents that users are most likely to request in advance according to their interests can achieve better performances.
However, it is difficult to predict which content users will like, as their interests are usually subjective and time-varying.
Besides, due to the limitation of network communication resources, e.g., bandwidth and power resources, they need to be allocated to each user adequately.
Therefore, how to properly design the communication-computing-caching (3C) policy to minimize the VR service delay as well as the energy consumption of the local device in the network deserves further discussion.

In this paper, an edge-terminal cooperative VR service system is proposed where the edge-cloud server is able to analysis the user interest by a proposed Bert model with the comments collected from the users' local devices.
With the predicted user interest, the communication, computing and caching policy is jointly designed for minimizing the weighted sum of VR service delay and energy consumption of local device with user fairness guaranteed.
The accuracy of the proposed method of interest analysis and the effectiveness of the proposed joint 3C optimization scheme are both validated by extensive simulations.
Main contributions are summarized as follows.

\begin{itemize}
\item \textit{A multiuser MEC-based mobile VR delivery framework is proposed to balance the VR service delay and energy consumption}:
    The framework for mobile VR delivery is proposed under four scenarios.
    Each scenario has a different VR service cost due to different computing and caching policy.
    In this framework, each raw VR content is allowed to be computed either on the edge-cloud server or local device.
    A portion of the raw VR content can be cached on local device in advance.

\item \textit{Applying sentiment analysis method to analysis users' interests in order to predict their request probability for all VR contents}:
    In the proposed framework, an advanced sentiment analysis method is applied to analysis the users' subjective interest among all contents with the given users' comments collected from all local devices.
    The proposed interest analysis method can be closer to the subjective feelings of users than other comparison schemes, which in turn can lead to a more accurate request probability matrix.
    Numerical results shows that the proposed sentiment analysis method to predict the users' request probability matrix is superior to the other three assumed probability distribution of user's requests which are commonly used in other papers.

\item \textit{A joint communication-computing-caching optimization policy is proposed to minimize the VR service cost}:
    The joint 3C problem is first decomposed into joint caching and computing subproblem and bandwidth allocation subproblem, and the joint 3C optimization policy can be obtained by solving the two subproblems separately by alternating iterations.
    Numerical results illustrate that the proposed joint 3C optimization policy performs better than the other three baseline schemes.

\end{itemize}

The remainder of this paper is as follows.
Related works are presented in Section 2.
Section 3 introduces the system model for VR services and formulates the problem that needs to be optimized.
Section 4 formulates the optimization problem and obtains the optimal policy named joint 3C optimization policy.
Simulation results are provided in Section 5.
Finally, Section 6 concludes the paper.

\section{Related works}

In the literature, many contributions have been devoted in mobile VR delivery architecture \cite{8728029,8713498,9693960,8664595}.
Reference \cite{8728029} propose a single-user MEC-based framework to minimize the average required transmission rate for VR content delivery, in which edge caching and edge computing are two key enabling technologies to improve the mobile VR delivery performance.
Similarly, a single-user fog radio access network is proposed to maximize the average tolerant delay for mobile VR delivery in \cite{8713498}.
In this network, the raw VR content can be cached in advance or computed on both fog access points and mobile VR devices to enhance the performance of VR service.
However, in single-user case, there is no need to consider the communication resource as there is only one user in the network.
In multi-user case, how to rationalise the allocation of communication resources available in the network to each user requires further consideration.
Reference \cite{9693960} adopts a NOMA-based MEC architecture which contains remote server, macro base station (MBS), small base station (SBS) and users, respectively, where each base station is equipped with an edge server.
Through the combination of the multicast transmission mode of MBS and the unicast transmission mode of SBS, the transcoding and delivery delay for VR service can be effectively reduced.
However, NOMA is generally suitable for transmitting small amounts of data in single-user scenario, and is not suitable for transmitting large amounts of data at high rates like such as VR contents due to interference among different users.
Reference \cite{8664595} propose a collaborative cloud and edge computing scheme to minimize the weighted-sum delay of all local devices in the network where tasks can be partially processed at the cloud and edge server.
However, in \cite{8664595}, it is assumed that each user has been associated with a corresponding base station which equipped with a MEC server in advance, and the communication resources of backhaul links between each device and its associated edge server equipped on the corresponding base station is assumed to be the same, which is unpractical.
The practical scenario is more likely to be a case where multiple local devices are jointly occupying limited communication resources within the same cell to perform a high-speed, high quality and large-data-volume VR services which worth further research.

Caching the raw VR content in advance according to user's interest can also improve the VR service performance in MEC-based network.
Hence, the request probability matrix for user to request those contents should first be determined before caching, which is a characterisation of the user's degree of interest in the corresponding content.
Most of the existing literature characterises the actual interest of users through an assumed distribution \cite{8728029,8713498,9693960,9013251,9893020,10062546,9839231,9843968}.
Reference \cite{8728029,8713498,9013251} consider that the probability of a user's request for all contents follows a uniform distribution, i.e., $P_i=1/N$ for each content $i\in \mathcal{N}$, $N=|\mathcal{N}|$.
Reference \cite{8728029,8713498,9693960,9893020,10062546,9839231,9843968} assume that the probability of a user's request for each content satisfies a Zipf distribution, i.e., the higher the popularity of the content, the higher the probability that it may be requested.
However, how the popularity of content is determined is not specified in detail.
In addition, those two aforementioned objective statistical distributions cannot accurately characterize the subjective interests of users in the network.
Therefore, how to accurately depict the probability matrix of requests for all contents by all users requires further consideration.

During the VR service, performance varies among users due to the different 3C resources allocated to each user.
In \cite{8728029}, optimization problem is formulated to minimize the average required transmission rate for VR service required by the user.
Reference \cite{8713498} propose an optimization problem to maximize the average tolerant delay.
The optimization problem posed in reference \cite{9693960} is to minimize the average delay which contains the preparation delay and the delivery delay.
Similarly, optimization problems are presented to minimize the average end-to-end delay for each user in \cite{8332500,9667509}.
However, the aforementioned literatures only optimize the average performance for all users.
This may result in a better performance for good performance user, but worse performance for poor counterpart.
The overall experience of VR services is dominated by the worst experience user \cite{9268953}, thus user fairness is also an important factor that should be taken into full consideration during the mobile VR delivery.

\section{System model}
\begin{figure}[h]
\centering
\includegraphics[width=1.0\linewidth]{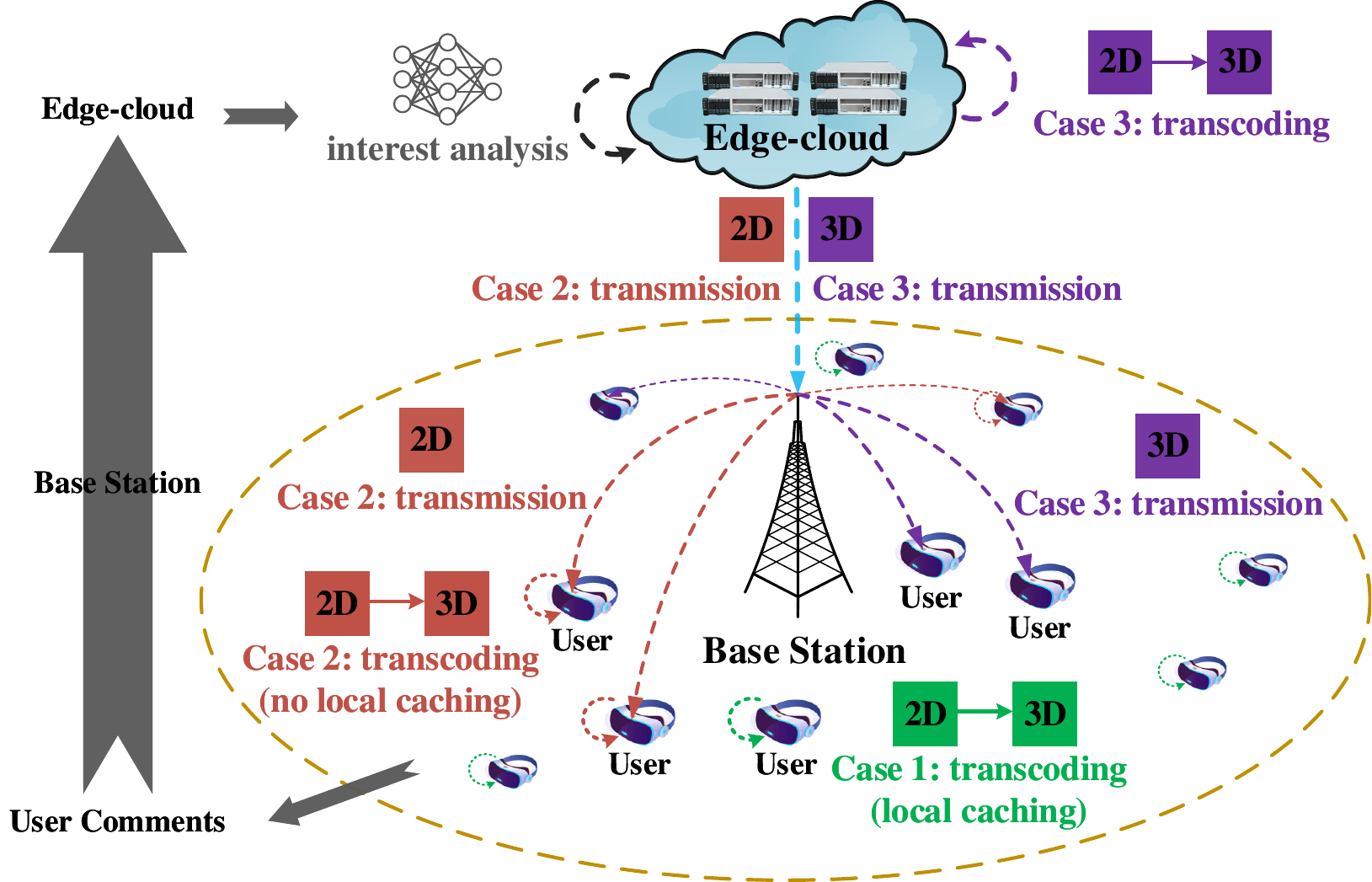}
\caption{Edge-terminal cooperative VR service delivery model.}
\label{fig1}
\end{figure}

As depicted in Fig. 1, the considered VR service delivery occurs within a small cell, which contains an edge-cloud server, a base station and several user equipments (UE).
In the cell, the edge-cloud server is configured with powerful computing and caching capabilities and the base station is responsible for delivering the pending and processed data to UE.
The edge-cloud server and the base station are connected by a wired link, so the communication time between the two is negligible.
Compared with the edge-cloud server, each UE has limited computing resource and caching capacity.

The whole VR video production steps include stitching, equirectangular projection, extraction, projection and rendering \cite{8728029}.
Among those steps, the first three steps need to transmit and process lots of data, so these works could be sent to remote cloud server in advance, which equipped with extremely strong computing and caching ability.
That projection can be properly allocated to the UE, whose ability counterpart is much weaker than cloud server, due to its lightweight computing requirements.
Therefore, in this work, our model only consider the projection procedure, which consists that change the two-dimensional (2D) monocular video chunks into the corresponding three-dimensional (3D) stereoscopic video chunks.

The overall VR service process can be referred to Fig. 1.
In specific, the 2D monocular video chunks should be calculated into the 3D stereoscopic video chunks counterpart on edge-cloud server or local UEs.
Only in this way can the UE continue the following and last step in VR video production, rendering, which guaranteeing successful watching VR videos by users.
After watching VR videos, users' comments on these videos will be fed back to the edge-cloud server.
These comments will be analyzed as guidelines for the edge-cloud server to decide which VR video chunks should be cached on each local UE during off-peak hours.

\subsection{VR service model}

Denote the set of VR video chunk contents by $\mathcal{I}$, and $N=\lvert\mathcal{I}\rvert$ is the number of contents in the library $\mathcal{I}$.
For each VR content $i\in\mathcal{I}\triangleq\{1,\cdots,N\}$, it can be characterized with a tuple $(D^{pl}, D_i^{sp}, \sigma_i)$ during the projection from 2D monocular video chunks to 3D counterpart, where $D^{pl}$ and $D_i^{sp}$ denote the size of 2D and 3D version (in bit), respectively, and $\sigma_i$ is the processing density (in CPU cycles per bit).
Typically, the size of 3D version should be at least twice larger than the 2D version, e.g., $D_i^{sp}/D^{pl}\geq2$.
For simplicity, this paper assumed that all 2D monocular video chunks have the same size.

Denote with $\mathcal{U}$ the set of users, whose local devices are denote as ${UE}_{1}, \cdots, {UE}_{u}, \cdots$, respectively, where $u\in\mathcal{U}\triangleq\{1,\cdots,M\}$ and $M=\lvert\mathcal{U}\rvert$ is the total number of these users.
Assuming that in the this cell, the channel allocated to each user does not overlap and can be accessed orthogonally.
Each UE is allocated a portion of the total bandwidth $B$ (in MHz) orthogonally to avoid interference between them.
Denote the normalization bandwidth allocated to each user $u$ by $a_u\in[0, 1]$, hence there exist $\sum_{u\in\mathcal{U}}a_u\leqslant1$.
For simplicity, the power allocated to each user $UE_u$, $P_u$ (in watt) , are assumed to be proportional to each user's bandwidth, e.g. $P_u/(a_uB)=pw$, $pw$ is a constant in this paper.
Therefore, the transmission rate $r_u$ between user $u$ and edge-cloud server can be obtained as:

\begin{equation}\label{equ-1}
r_u=a_uBlog_2{(1+\frac{pwh_u}{N_0})}
\end{equation}
where the channel (from edge-cloud server to $UE_u$) gain and noise are donated with $h_u$, and $N_0$, respectively.

\subsection{Cost under different scenarios}
Each content $i\in\mathcal{I}$ can be cached on either edge-cloud side or local side.
Assuming that all of the 2D monocular video contents have been cached on edge-cloud server in advance.
Denote with $c_{ui}\in\{0, 1\}$ the cache policy, where $c_{ui}=1$ represent content $i$ has been already cached at $UE_u$, and $c_{ui}=0$ otherwise.
Similarly, the computation policy can be donated by $d_{ui}\in\{0, 1\}$, where $d_{ui}=1$ donates that content $i$ is transcoded at $UE_u$, and $d_{ui}=0$ otherwise.
In what follows, the VR service cost is analyzed according to different values of $c_{ui}$ and $d_{ui}$ in four different combination cases.

\begin{itemize}
    \item $c_{ui}=1$ \& $d_{ui}=1$.

    In this case, content $i$ is cached and transcoded at $UE_u$.
    Hence, the VR service time consumption only contains local transcoding delay.
    Denote with $f_u^{loc}$, $p_u^{loc}$ the local computing capability (in CPU cycles/s) and the power consumption of $UE_u$ under local executing state (in watt), respectively.
    Therefore, the corresponding VR service delay and local energy consumption for user $UE_u$ to request content $i$ are formulated as follows:

    \begin{equation}\label{equ-2}
    T_{ui}^{loc1}=\frac{D^{pl}\sigma_i}{f_u^{loc}}
    \end{equation}

    \begin{equation}\label{equ-3}
    E_{ui}^{loc1}=p_u^{loc}\frac{D^{pl}\sigma_i}{f_u^{loc}}
    \end{equation}

    \item $c_{ui}=0$ \& $d_{ui}=1$.

    In this case, the input data, e.g. 2D version of content $i$, are not cached at local side.
    The content $i$ is cached at the edge-cloud server side and transcoded at local side through transmitting that content from edge-cloud server to local side.
    Denote with $p_u^{com}$ the power consumption of $UE_u$ under transmission state (in watt).
    Hence, the VR service time consumption contains the transmission delay and the local transcoding delay, and those two types of consumption can be depicted as:

    \begin{equation}\label{equ-4}
    T_{ui}^{loc2}=\frac{D^{pl}}{r_u}+\frac{D^{pl}\sigma_i}{f_u^{loc}}
    \end{equation}

    \begin{equation}\label{equ-5}
    E_{ui}^{loc2}=p_u^{com}\frac{D^{pl}}{r_u}+p_u^{loc}\frac{D^{pl}\sigma_i}{f_u^{loc}}
    \end{equation}

\end{itemize}

Combining (\ref{equ-2})-(\ref{equ-5}), when content $i$ is transcoded at $UE_u$, there holds:

\begin{equation}\label{equ-6}
T_{ui}^{loc}=T_{ui}^{loc1}c_{ui}+T_{ui}^{loc2}(1-c_{ui})
\end{equation}

\begin{equation}\label{equ-7}
E_{ui}^{loc}=E_{ui}^{loc1}c_{ui}+E_{ui}^{loc2}(1-c_{ui})
\end{equation}

\begin{itemize}
    \item $c_{ui}=0$ \& $d_{ui}=0$.

    In this case, content $i$ is cached and transcoded at edge-cloud server.
    Hence, the VR service time consumption contains the transcoding delay at the edge-cloud server and the transmission delay from edge-cloud server to local device.
    Denote with $f^{c}$, $p_u^{id}$ the edge-cloud server computing capability (in CPU cycles per second) and the power consumption of $UE_u$ under local idle state (in watt), respectively.
    Therefore, those two kinds of consumption can be depicted as:

    \begin{equation}\label{equ-8}
    T_{ui}^c=\frac{D^{pl}\sigma_i}{f^c}+\frac{D_i^{sp}}{r_u}
    \end{equation}

    \begin{equation}\label{equ-9}
    E_{ui}^c=p_u^{id}\frac{D^{pl}\sigma_i}{f^c}+p_u^{com}\frac{D_i^{sp}}{r_u}
    \end{equation}

    \item $c_{ui}=1$ \& $d_{ui}=0$.

    In this case, content $i$ is cached at local side and transcoded at edge-cloud server.
    The input data required for calculation process on the edge-cloud server can be directly obtained by edge-cloud server itself, rather than local side.
    Hence, caching contents on the $UE_u$ is a waste of local caching resource, and cannot reduce the delay and energy consumption any more.
    This case is no longer need to be considered.

\end{itemize}

According to equation (\ref{equ-6})-(\ref{equ-9}), the final delay and energy consumption holds:

\begin{equation}\label{equ-10}
\begin{aligned}
T_{ui}&=T_{ui}^{loc}d_{ui}+T_{ui}^c(1-d_{ui})\\
&=(\frac{D^{pl}}{r_u}(1-c_{ui})+\frac{D^{pl}\sigma_i}{f_u^{loc}}c_{ui})d_{ui}+(\frac{D^{pl}\sigma_i}{f^c}+\frac{D_i^{sp}}{r_u})(1-d_{ui})
\end{aligned}
\end{equation}

\begin{equation}\label{equ-11}
\begin{aligned}
E_{ui}&=E_{ui}^{loc}d_{ui}+E_{ui}^c(1-d_{ui})\\
&=(p_u^{com}\frac{D^{pl}}{r_u}(1-c_{ui})+p_u^{loc}\frac{D^{pl}\sigma_i}{f_u^{loc}}c_{ui})d_{ui}\\
&\quad+(p_u^{id}\frac{D^{pl}\sigma_i}{f^c}+p_u^{com}\frac{D_i^{sp}}{r_u})(1-d_{ui})
\end{aligned}
\end{equation}

\textit{\textbf{Lemma~1:}} For any user $u\in\mathcal{U}$ and content $i\in\mathcal{I}$, there always holds $c_{ui}d_{ui}=c_{ui}$.

\textit{\textbf{Proof:}} As mentioned in Case 4, in the practical network, if there is no request for transcoding content from 2D version to 3D version on the local service, there is no need for such content to be cached locally.
Therefore, there always holds $c_{ui}\leq d_{ui}$.
Further, $c_{ui}d_{ui}=c_{ui}$ holds due to $c_{ui}\in\{0, 1\}$ and $d_{ui}\in\{0, 1\}$.
$\hfill\blacksquare$

According to \textit{\textbf{Lemma~1}}, equation (\ref{equ-10}) and (\ref{equ-11}) can be reduced to:

\begin{equation}\label{equ-12}
\begin{aligned}
T_{ui}&=[\frac{D^{pl}}{r_u}(1-c_{ui})+\frac{D^{pl}\sigma_i}{f_u^{loc}}]d_{ui}+(\frac{D^{pl}\sigma_i}{f^c}+\frac{D_i^{sp}}{r_u})(1-d_{ui})\\
&=(\frac{D^{pl}\sigma_i}{f^c}+\frac{D_i^{sp}}{r_u})-\frac{D^{pl}}{r_u}c_{ui}
+(\frac{D^{pl}}{r_u}+\frac{D^{pl}\sigma_i}{f_u^{loc}}-\frac{D^{pl}\sigma_i}{f^c}-\frac{D_i^{sp}}{r_u})d_{ui}\\
\end{aligned}
\end{equation}

\begin{equation}\label{equ-13}
\begin{aligned}
E_{ui}&=[p_u^{com}\frac{D^{pl}}{r_u}(1-c_{ui})+p_u^{loc}\frac{D^{pl}\sigma_i}{f_u^{loc}}]d_{ui}\\
&\quad+(p_u^{id}\frac{D^{pl}\sigma_i}{f^c}+p_u^{com}\frac{D_i^{sp}}{r_u})(1-d_{ui})\\
&=(p_u^{id}\frac{D^{pl}\sigma_i}{f^c}+p_u^{com}\frac{D_i^{sp}}{r_u})-p_u^{com}\frac{D^{pl}}{r_u}c_{ui}\\
&\quad+(p_u^{com}\frac{D^{pl}}{r_u}+p_u^{loc}\frac{D^{pl}\sigma_i}{f_u^{loc}}
-p_u^{id}\frac{D^{pl}\sigma_i}{f^c}-p_u^{com}\frac{D_i^{sp}}{r_u})d_{ui}\\
\end{aligned}
\end{equation}

\subsection{Problem Formulation}
Due to the difference in VR videos among users, different users should have different probabilities to request for those contents.
Denote $\zeta_{ui}\in[0,1]$ as the probability for $UE_u$ to request for content $i$.
Thus, the expectation of delay and energy consumption for $UE_u$ holds:

\begin{equation}\label{equ-14}
T_u=\sum_{i=1}^{N}{\zeta_{ui}T_{ui}}
\end{equation}

\begin{equation}\label{equ-15}
E_u=\sum_{i=1}^{N}{\zeta_{ui}E_{ui}}
\end{equation}

In this paper, the cost of $UE_u$ is defined as the weighted sum of delay and energy consumption as $Cost_u=\lambda^eE_u+\lambda^tT_u$, where $\lambda^t,\lambda^e\in[0,1]$, denote the weights of delay and energy consumption, respectively.
This paper aims to minimize the maximum cost among all $UE$s under the cache and communication constraints as follows:

\begin{equation}\label{equ-16}
\begin{aligned}
&(\text{P1})\quad\underset{\{c_{ui}\},\{a_{u}\}}{\text{min}}\,\underset{u\in\mathcal{U}}{\text{max}}\quad Cost_u\\
&\quad\quad\quad\text {s.t.} \quad \text{(C1)}:~\sum_{i=1}^N c_{ui} \leq C_u \\
&\quad \quad \quad \quad \quad \,\text{(C2)}:~0 \leq a_u \leq 1 \\
&\quad \quad \quad \quad \quad \,\text{(C3)}:~\sum_{u=1}^M a_u \leq 1 \\
\end{aligned}
\end{equation}
Where constraint (C1) indicates that the number of local cached contents should not exceed the local cache capability, $C_u$, for each device.
Constraint (C2) and (C3) means that the bandwidth allocated to each UE is nonnegative and the sum of them should not exceed the total available bandwidth value.
Problem $(\text{P1})$ is a non-convex optimization problem due to its minimize-maximum formulation, which cannot be solved by traditional convex tools directly.
Moreover, as content request probability $\{\zeta_{ui}\}$ depends on user interest, it cannot be characterized with explicit expression, which increases the solving difficulty of problem $(\text{P1})$.

\section{VR service policy design}

In this section, a VR service policy is designed to minimize the service cost with consideration of performance fairness guarantee among users.
The optimization parameters in problem $(\text{P1})$ include cache policy $\{c_{ui}\}$, computation policy $\{d_{ui}\}$ and communication policy $\{a_u\}$, in which the cache and computation policy is independent with the communication policy according to the optimization goal and constraints in $(\text{P1})$.
But before the optimization, the users' request probability should be acquired due to equation (\ref{equ-14}) and (\ref{equ-15}).
Therefore, problem $(\text{P1})$ can be decoupled into a request probability solving subproblem for each user, a joint caching and computing subproblem for each user and a bandwidth allocation subproblem for all users, respectively.
Those three subproblems will be illustrated separately in the following.

\subsection{Request probability solving subproblem}

As mentioned above, the request probability solving subproblem should be managed firstly.
In other works, the conditions that users request for contents are not be carefully considered.
Usually, the probability of user requests for contents are assumed to follow the uniform distribution ($P\propto 1/{N}$) or Zipf distribution ($P\propto 1/{i^\gamma}$).
However, the above distributions cannot represent those users' real subjective conditions.
Therefore, in this work, sentiment analysis of user's comments is used as a guideline for users' request probability matrix's acquisition.

As depicted in Fig. 1., the users' comments will be firstly collected and transmitted to the edge-cloud server, in which the comments will be fully analyzed by the deep learning neural networks.
Then, the results of the sentiment analysis will serve as the basis for obtaining the user's request probability matrix.
Due to the assumption that the action of content caching can be done during the local device's idle time, there is no need to consider the computational speed of those neural networks.
Thus which deep learning neural network to choose only depends on the accuracy of their sentiment analysis results.

The common methods for analyzing the text mainly include TextCNN, TextRNN, FastText, Bert and so on.
TextCNN is suitable for short text analysis, while TextRNN can further consider the correlation between words that are far apart in one sentence.
Compared with the first two methods, Fasttext can achieve a faster computation speed at the cost of accuracy.
Bert enable each word to interact with others by using the self attention mechanism in Transformer \cite{vaswani2017attention}, thereby capturing richer contextual information and achieving higher accuracy, though it may cost more computation time.
Further experimental results are shown in $\text{Table~\ref{table1}}$ based on the IMDB dataset\cite{maas2011learning}, where the number of samples in the training set, validation set and test set are 70000, 15000 and 15000 respectively.

\begin{table}[h]\footnotesize
\centering
\begin{threeparttable}
\setlength{\abovecaptionskip}{0pt}
\caption{Experiment on accuracy among different methods.}
\captionsetup{justification = centerlast}
\label{table1}
\begin{tabular}{|p{2.0cm}<{\centering} | p{1.5cm}<{\centering} | p{1.5cm}<{\centering} | p{1.5cm}<{\centering}|}
\hline
\multicolumn{1}{|c|}{\multirow{2}{*}{Method}} & \multicolumn{3}{c|}{Accuracy}\\ \cline{2-4}
 & train\_acc  & val\_acc & test\_acc \\ \hline
 TextCNN\cite{kim2014convolutional} & $\textbf{94.17\%}$ & 87.26\%  & 85.89\%  \\ \hline
 TextRNN\cite{liu2016recurrent} & 86.66\% & 88.79\% & 87.85\%  \\ \hline
 FastText\cite{joulin2016bag} & 86.87\% & 85.66\% & 85.45\%  \\ \hline
 Bert\cite{devlin2018bert} & 93.07\% & $\textbf{91.23\%}$ & $\textbf{91.87\%}$  \\ \hline
\end{tabular}
\end{threeparttable}
\end{table}

As mentioned earlier, only the accuracy of sentiment analysis should be considered in this paper instead of the computation time due to the idle time transmission policy.
According to both the simple theoretical analysis and rigorous experiments on those methods above, Bert is the most suitable method for sentiment analysis in this paper.
Thus, Bert is chose to acquire the users' request probability matrix to solve the request probability subproblem.
Specific steps are as follows:

\begin{itemize}

\item Firstly, the Bert model is used to analysis the sentiment of the collected comments from each user on each content and to derive the sentiment value of the comments on the corresponding content.
      The sentiment value is between 0 and 1, the closer it is to 1, the more interested the user is in the corresponding content and vice versa.

\item Subsequently, normalize each user's sentiment values for all content so that the sum of them equals 1.
     After normalization, the sentiment value of each user for each content can be used as the request probability, thus obtaining the request probability matrix.

\end{itemize}

\subsection{Joint caching and computing subproblem}

With given optimal bandwidth allocation policy $\{a_u\}$, the transmission rate $\{r_u^*\}$ of all users is transformed into constant, according to equation (\ref{equ-1}).
Hence, problem $(\text{P1})$ can be further reduced to problem $(\text{P2})$, which is displayed as follows:

\begin{equation}\label{equ-17}
\begin{aligned}
&(\text{P2})\quad\underset{\{c_{ui}\}}{\text{min}}\,\underset{u\in\mathcal{U}}{\text{max}}\quad Cost_u\\
&\quad\quad\quad\text {s.t.} \quad \text{(C1)}:~\sum_{i=1}^N c_{ui} \leq C_u \\
\end{aligned}
\end{equation}

Among all users, denote $UE_{u^{\prime}}$ as the user who reach the maximum cost value, e.g., $Cost_{u^{\prime}} = \text{max}(Cost_1, Cost_2, \cdots, Cost_M)$, $u^{\prime}\in\mathcal{U}$.
Then problem $(\text{P2})$ can be transformed into problem $(\text{P3})$:

\begin{equation}\label{equ-18}
\begin{aligned}
&(\text{P3})\quad\underset{\{c_{u^{\prime}i}\}}{\text{min}}\quad Cost_{u^{\prime}}\\
&\quad\quad\quad\text {s.t.} \quad (\text{C1}^{\prime}):~\sum_{i=1}^N c_{u^{\prime}i} \leq C_{u^{\prime}} \\
\end{aligned}
\end{equation}
where

\begin{equation}\label{equ-19}
\begin{aligned}
Cost_{u^{\prime}}
&=\lambda^eE_{u^{\prime}}+\lambda^tT_{u^{\prime}}=\sum_{i=1}^N{\lambda^eE_{u^{\prime}i}+\lambda^tT_{u^{\prime}i}}\\
&=\sum_{i=1}^N{\zeta_{u^{\prime}i}[(\lambda^ep_{u^{\prime}}^{id}+\lambda^t)\frac{D^{pl}\sigma_i}{f^c}+(\lambda^ep_{u^{\prime}}^{com}+\lambda^t)\frac{D_i^{sp}}{r_u^{\prime}}]}\\
&\quad-\sum_{i=1}^N{\zeta_{u^{\prime}i}(\lambda^ep_{u^{\prime}}^{com}+\lambda^t)\frac{D^{pl}}{r_{u^{\prime}}}c_{u^{\prime}i}}\\
&\quad+\sum_{i=1}^N{\zeta_{u^{\prime}i}[(\lambda^ep_{u^{\prime}}^{com}+\lambda^t)\frac{D^{pl}}{r_{u^{\prime}}}+(\lambda^ep_{u^{\prime}}^{loc}+\lambda^t)\frac{D^{pl}\sigma_i}{f_{u^{\prime}}^{loc}}}\\
&\quad{-(\lambda^ep_{u^{\prime}}^{id}+\lambda^t)\frac{D^{pl}\sigma_i}{f^c}-(\lambda^ep_{u^{\prime}}^{com}+\lambda^t)\frac{D_i^{sp}}{r_{u^{\prime}}}]d_{u^{\prime}i}}\\
\end{aligned}
\end{equation}

For convenience, we use $A_{u^{\prime}i}$, $B_{u^{\prime}i}$, $F_{u^{\prime}i}$, $G_{u^{\prime}i}$ and $H_{u^{\prime}i}$ to replace the corresponding five parts in the RHS of equation (\ref{equ-19}):
$$A_{u^{\prime}i}=\zeta_{u^{\prime}i}(\lambda^ep_{u^{\prime}}^{com}+\lambda^t)\frac{D^{pl}}{r_{u^{\prime}}}$$
$$B_{u^{\prime}i}=\zeta_{u^{\prime}i}(\lambda^ep_{u^{\prime}}^{loc}+\lambda^t)\frac{D^{pl}\sigma_i}{f_{u^{\prime}}^{loc}}$$
$$F_{u^{\prime}i}=\zeta_{u^{\prime}i}(\lambda^ep_{u^{\prime}}^{id}+\lambda^t)\frac{D^{pl}\sigma_i}{f^{c}}$$
$$G_{u^{\prime}i}=\zeta_{u^{\prime}i}(\lambda^ep_{u^{\prime}}^{com}+\lambda^t)\frac{D_i^{sp}}{r_{u^{\prime}}}$$
$$H_{u^{\prime}}=\sum_{i=1}^N{\zeta_{u^{\prime}i}[(\lambda^ep_{u^{\prime}}^{id}+\lambda^t)\frac{D^{pl}\sigma_i}{f^c}}
+(\lambda^ep_{u^{\prime}}^{com}+\lambda^t)\frac{D_i^{sp}}{r_{u^{\prime}}}]$$

Then equation (\ref{equ-19}) can be further reduced to:

\begin{equation}\label{equ-20}
\begin{aligned}
Cost_{u^{\prime}}&=-\sum_{i=1}^N{A_{u^{\prime}i}}c_{u^{\prime}i}
+\sum_{i=1}^N{(A_{u^{\prime}i}+B_{u^{\prime}i}-F_{u^{\prime}i}-G_{u^{\prime}i})}d_{u^{\prime}i}
+H_{u^{\prime}}\\
&=\sum_{i=1}^N{A_{u^{\prime}i}}(d_{u^{\prime}i}-c_{u^{\prime}i})
+\sum_{i=1}^N{(B_{u^{\prime}i}-F_{u^{\prime}i}-G_{u^{\prime}i})}d_{u^{\prime}i}+H_{u^{\prime}}
\end{aligned}
\end{equation}

Observed from equation (\ref{equ-20}), the optimal caching policy $\{c_{u^{\prime}i}\}$, and computing policy $\{d_{u^{\prime}i}\}$ can be obtained by analyzing the coefficient of $c_{u^{\prime}i}$ and $d_{u^{\prime}i}$.
First, it is obviously that all of the $A_{u^{\prime}i}$, $B_{u^{\prime}i}$, $F_{u^{\prime}i}$, $G_{u^{\prime}i}$, $H_{u^{\prime}}$ are non-negative.
In the practical network, the power consumption of $UE_u$ under local executing rate is greater than the transmission rate counterpart, which is greater than the idle rate, e.g., $p_u^{loc}>p_u^{com}>p_u^{id}$.
Next, according to the four cases described in Section 3.2, the first item in $Cost_{u^{\prime}}$ is non-negative due to $c_{u^{\prime}i}\leq d_{u^{\prime}i}$.
Then, for the second term, based on the coefficient of $d_{u^{\prime}i}$, it can be divided into the following two situations:

\text{\bf{Case 1:}}\quad$A_{u^{\prime}i}+B_{u^{\prime}i}-F_{u^{\prime}i}-G_{u^{\prime}i} > 0$.

The minimize value of $Cost_{u^{\prime}}$ can be obtained when $d_{u^{\prime}i}=0$.
Hence, $c_{u^{\prime}i}=0$ due to $c_{u^{\prime}i}\leq d_{u^{\prime}i}$.
In this case, the contents are cached and computed at the edge server.

\text{\bf{Case 2:}}\quad$A_{u^{\prime}i}+B_{u^{\prime}i}-F_{u^{\prime}i}-G_{u^{\prime}i} \leq 0$.

The minimize value of $Cost_{u^{\prime}}$ should be acquired at $d_{u^{\prime}i}=1$.
Under the conditions of $d_{u^{\prime}i}=1$, the value of $Cost_{u^{\prime}}$ at $c_{u^{\prime}i}=1$ is smaller than $c_{u^{\prime}i}=0$ counterpart.
However, constraint $(\text{C1}^{\prime})$ limits the number of contents whose $c_{u^{\prime}i}$ reach 1.
Therefore, the cache capacity can be fully used by arranging the largest first $C_u^{\prime}$ contents's $c_{u^{\prime}i}$ to 1, while others 0.
In this case, the processing task is accomplished at the local side, and the content are distributed on edge-cloud server and local side according to the optimal policy.
Specifically, denote set $\mathcal{I}_1$ as the set of contents satisfying Case 2.
Sort all of the elements in $A_{u^{\prime}i}$ in descending order.
Pick the first $\text{min}\{C_{u^{\prime}}, \lvert {\mathcal{I}_1} \rvert\}$ content and cache them on $UE_{u^{\prime}}$, e.g., $c_{u^{\prime}i}=1$, otherwise $c_{u^{\prime}i}=0$.
In this way can problem $(\text{P3})$ be solved, the detailed algorithm process is shown in $
\text{Algorithm~1}$.

\begin {algorithm}[h] \footnotesize
	\caption{Caching and Computing Policy Design}
    \label{alg1:}
	\textbf{Input:} $i\in\{1,\cdots,N\}$, $u\in\{1,\cdots,M\}$, ${\bf{\zeta}}=[\zeta_{ui}]_{M\times N}$, $\textbf{a}_u=[a_{u}]_{M\times 1}$.\\
	\textbf{Output:} $\textbf{C}=[c_{ui}]_{M\times N}$, $\textbf{D}=[d_{ui}]_{M\times N}$.
	\begin{algorithmic}[1]
        \State Initialize $\textbf{C}=\textbf{D}=\textbf{0}_{M\times N}$.
        \State Calculate $\textbf{A}=[A_{ui}]_{M\times N}$, $\textbf{B}=[B_{ui}]_{M\times N}$, $\textbf{F}=[F_{ui}]_{M\times N}$, $\textbf{G}=[G_{ui}]_{M\times N}$, where $A_{ui}=\zeta_{ui}(\lambda_{u}^ep_{u}^{com}+\lambda_{u}^t){D^{pl}}/{r_{u}}$,
        $B_{ui}=\zeta_{ui}(\lambda_{u}^ep_{u}^{loc}+\lambda_{u}^t){D^{pl}\sigma_i}/{f_{u}^{loc}}$,
        $F_{ui}=\zeta_{ui}(\lambda_{u}^ep_{u}^{id}+\lambda_{u}^t){D^{pl}\sigma_i}/{f^{c}}$, $G_{ui}=\zeta_{ui}(\lambda_{u}^ep_{u}^{com}+\lambda_{u}^t){D_i^{sp}}/{r_{u}}$.
		\State Set $\mathbf{\Delta}=\textbf{A}=[A_{ui}]_{M\times N}$.
        \If {$A_{ui}+B_{ui}-F_{ui}-G_{ui} \geq 0$}
            \State \textbf{set} $\mathbf{\Delta}_{ui}=-1$.
        \Else
            \State \textbf{set} $d_{ui}=1$.
        \EndIf
        \For {$u=1,\cdots,M$}
            \State Sort the elements in $\Delta_u$ in descending order as $\Delta_u^{\prime}$.
            \State Mark the first $C_u$ elements greater than $0$ in $\Delta_u^{\prime}$, find the corresponding positions in $\Delta_u$ and set the corresponding positions in $[c_u]$ to 1.
        \EndFor
	\end{algorithmic}
\end{algorithm}

\subsection{Bandwidth allocation subproblem}

With given users' request probability matrix $\bf{\zeta}=[\zeta_{ui}]_{M\times N}$,
optimal caching policy $\textbf{C}^*=[c_{ui}^*]_{M\times N}$ and computing policy $\textbf{D}^*=[d_{ui}^*]_{M\times N}$, problem $(\text{P1})$ can be further reduced to problem $(\text{P4})$, which is displayed as follows:

\begin{equation}\label{equ-21}
\begin{aligned}
&(\text{P4})\quad\underset{\{a_{u}\}}{\text{min}}\,\underset{u\in\mathcal{U}}{\text{max}}\quad Cost_u\\
&\quad\quad\quad\text {s.t.} \quad \text{(C1)}:~0 \leq a_u \leq 1 \\
&\quad \quad \quad \quad \quad \,\text{(C2)}:~\sum_{u=1}^M a_u \leq 1 \\
\end{aligned}
\end{equation}

\textit{\textbf{Lemma~2:}} For any two different users $u_1,u_2\in\mathcal{U}$ under given request probability ${\bf{\zeta}}=[\zeta_{ui}]_{M\times N}$, optimal caching policy $\textbf{C}^*=[c_{ui}^*]_{M\times N}$ and computing policy $\textbf{D}^*=[d_{ui}^*]_{M\times N}$, when the optimal bandwidth allocation $\textbf{a}_u^*=[a_u]_{M\times 1}$ is obtained, there holds:

\begin{equation}\label{equ-22}
Cost_{u_1}=Cost_{u_2}
\end{equation}

\textit{\textbf{Proof:}} As $\bf{\zeta}=[\zeta_{ui}]_{M\times N}$, $\textbf{C}^*=[c_{ui}^*]_{M\times N}$, $\textbf{D}^*=[d_{ui}^*]_{M\times N}$ are given, $\zeta_{ui}$, $c_{ui}$, $d_{ui}$ are constant in equation (\ref{equ-19}).
Then, for any $UE_u$, $\sum_{i=1}^N{\zeta_{ui}(\lambda^ep_{u}^{id}+\lambda^t){D^{pl}\sigma_i}/{f^{c}}}$,
$\sum_{i=1}^N{\zeta_{ui}(\lambda^ep_{u}^{loc}+\lambda^t){D^{pl}\sigma_i}d_{ui}/{f_u^{loc}}}$, $\sum_{i=1}^N{\zeta_{ui}(\lambda^ep_{u}^{id}+\lambda^t){D^{pl}\sigma_i}d_{ui}/{f^{c}}}$,
are constant, respectively.
Denote $const$ as the sum of these three items above, so that $Cost_u$ can be further transformed into:

\begin{equation}\label{equ-23}
\begin{aligned}
Cost_u&=\sum_{i=1}^N{\zeta_{ui}(\lambda^ep_{u}^{com}+\lambda^t)\frac{D_i^{sp}}{r_u}}
-\sum_{i=1}^N{\zeta_{ui}(\lambda^ep_{u}^{com}+\lambda^t)\frac{D^{pl}}{r_u}c_{ui}}\\
&\quad+\sum_{i=1}^N{\zeta_{ui}[(\lambda^ep_{u}^{com}+\lambda^t)\frac{D^{pl}}{r_u}-(\lambda^ep_{u}^{com}+\lambda^t)\frac{D_i^{sp}}{r_u}]d_{ui}}+const\\
&=\sum_{i=1}^N{\zeta_{ui}(\lambda^ep_{u}^{com}+\lambda^t)D^{pl}(d_{ui}-c_{ui})\frac{1}{r_u}}\\
&\quad+\sum_{i=1}^N{\zeta_{ui}(\lambda^ep_{u}^{com}+\lambda^t)D_i^{sp}(1-d_{ui})\frac{1}{r_u}}+const\\
\end{aligned}
\end{equation}

Due to the fact that $\zeta_{ui}, \lambda^e, \lambda^t \geq 0$, $0\leq c_{ui} \leq d_{ui} \leq 1$, $D_i^{sp} > D^{pl} >0$, the coefficient of $1/r_u$ is positive in $Cost_u$, e.g., $Cost_u\propto 1/r_u$.
According to equation (\ref{equ-1}) that $r_u\propto a_u$, there holds $Cost_u\propto 1/a_u$.

Next, \textit{\textbf{Lemma~2}} can be proved with the contradiction method.
Assuming that the optimal bandwidth allocation $\textbf{a}_u^*=[a_u]_{M\times 1}$ is acquired, $UE_1$ always suffers the maximum cost value among all users.
Therefore, there holds $Cost_{u_1}>Cost_{u_2}$.
Let $a_{u_1}^{\prime}=a_{u_1}+\Delta a$, $a_{u_2}^{\prime}=a_{u_2}-\Delta a$ where $\Delta a\rightarrow 0$.
In this way, the sum bandwidth value of all users still remains unchanged.
The new cost value of $UE_1$, $UE_2$, e.g., $Cost_{u_1}^{\prime}$, $Cost_{u_2}^{\prime}$, can be obtained according to the equation (\ref{equ-23}).
Since $Cost_u \propto 1/a_u$, there holds:

\begin{equation}\label{equ-24}
Cost_{u_1}>Cost_{u_1}^{\prime}>Cost_{u_2}^{\prime}>Cost_{u_2}
\end{equation}

Therefore, when other user's cost value remains unchanged, the new cost value of $UE_1$ becomes lower than before, which is contradictory with the assumption that $UE_1$ can always reach the maximum value among all users.
Thus, the cost value among all users are equal when the optimal bandwidth allocation policy $\textbf{a}_u^*=[a_u]_{M\times 1}$ is given.
$\hfill\blacksquare$

In order to minimize $\text{max}(Cost_1,\cdots,Cost_{u},\cdots,Cost_M)$, the bandwidth allocated to each user should be as large as possible, since $Cost_u \propto 1/a_u$.
Thus, when problem $(\text{P4})$ is solved, the cost value of each user should be equal, e.g., $Cost_1=\cdots=Cost_u=\cdots=Cost_M$, according to \textit{\textbf{Lemma~2}}.
At the same time, the sum value of all users' allocated bandwidth should reach the upper limit value, 1, of the constraint $(\text{C2})$, e.g., $\sum_{u=1}^M{a_u}=1$.
Hence, problem $(\text{P4})$ can be solved by bisection method, which is summarized in $\text{Algorithm~2}$.

\begin {algorithm}[h] \footnotesize
	\caption{Bandwidth Allocation Policy Design}
    \label{alg2:}
	\textbf{Input:} $i\in\{1,\cdots,N\}$, $u\in\{1,\cdots,M\}$, ${\bf\zeta}=[\zeta_{ui}]_{M\times N}$, $\textbf{C}=[c_{ui}]_{M\times N}$, $\textbf{D}=[d_{ui}]_{M\times N}$.\\
	\textbf{Output:} $\textbf{a}_u=[a_{u}]_{M\times 1}$, $Cost^{\text{max}}$.
	\begin{algorithmic}[1]
        \State Initialize $a^{\text{min}}=0$, $a^{\text{max}}=1$, $a_1=a_2=\cdots=a_M=1/M$.
        \Repeat
        \State Calculate ${\bf{Cost}}=[Cost_u]_{M\times 1}$
        \State Find the maximum element $Cost_{u^{\prime}}$ in $\bf{Cost}$ and return the corresponding index $u^{\prime}$.
        \State Set $Cost^{\text{max}}=Cost_{u^{\prime}}$.
        \State Set ${\bf{Cost}}=Cost^{\text{max}}\times \mathbf{1}_{M\times 1}$ and calculate the new bandwidth distribution $\mathbf{a}_{u^{\prime}}=[a_u^{\prime}]_{M\times 1}$ according to equation (\ref{equ-24}).
        \If {$\sum_{u=1}^M{a_u}<1$}
            \State Set $a^{\text{min}}=a_{u^{\prime}}^{\prime}$.
        \ElsIf {$\sum_{u=1}^M{a_u}>1$}
            \State Set $a^{\text{max}}=a_{u^{\prime}}^{\prime}$.
        \Else
            \State Jump out of the repeat loop.
        \EndIf
        \State Set $a_u^{\prime}=(a^{\text{max}}+a^{\text{min}})/2$.
        \Until{$\sum_{u=1}^M{a_u}=1$}
	\end{algorithmic}
\end{algorithm}

\subsection{Joint 3C optimization policy}

To solve problem $(\text{P1})$, the request probability matrix should be obtained firstly through analyzing different user's comments on different contents in order to get their subjective interests on those VR videos.
There interests can be served as the guideline for obtaining the request probability matrix $\bf{\zeta}$.
In this way can the request probability get much closer to the users' subjective ideas.

Next, the optimal VR service policy can be acquired by solving problem $(\text{P3})$ and $(\text{P4})$ alternatively.
Without loss of generality, the bandwidth allocation policy for each user is initialized as that all users share the total bandwidth resources.
Under this circumstances, the optimal caching policy $\textbf{C}^*=[c_{ui}^*]_{M\times N}$ and computing policy $\textbf{D}^*=[d_{ui}^*]_{M\times N}$ can be obtained by solving problem $(\text{P3})$ with $\text{Algorithm~1}$.
With given these two policies, the optimal bandwidth allocation policy $\textbf{a}_u^*=[a_u]_{M\times 1}$ can be obtained by solving problem $(\text{P4})$ with $\text{Algorithm~2}$.
At the same time, the former bandwidth allocation policy will be replaced by the latter.
Thus problem $(\text{P1})$ can be solved by iterating $\text{Algorithm~1}$ and $\text{Algorithm~2}$ alternatively until the value of $Cost^{\text{max}}$ is equal to its previous iteration.
The specific steps to solve problem $(\text{P1})$ are summarized in $\text{Algorithm~3}$ as follows.

\begin {algorithm}[h] \footnotesize
	\caption{Joint 3C Optimization Policy Design}
    \label{alg3:}
	\textbf{Input:} $i\in\{1,\cdots,N\}$, $u\in\{1,\cdots,M\}$, ${\bf{\zeta}}=[\zeta_{ui}]_{M\times N}$.\\
	\textbf{Output:} $\textbf{C}=[c_{ui}]_{M\times N}$, $\textbf{D}=[d_{ui}]_{M\times N}$, $\textbf{a}_u=[a_{u}]_{M\times 1}$, $Cost^{\text{opt}}$.
	\begin{algorithmic}[1]
        \State Initialize $\mathbf{a}_u={\bf{1}}_{M\times 1}/M$, $Cost^{\text{max}}=+\infty$, $Cost^{\text{opt}}=-1$.
        \While {$Cost^{\text{opt}}\neq Cost^{\text{max}}$}
        \State Set $Cost^{\text{opt}}= Cost^{\text{max}}$
        \State Find the optimal caching policy $\textbf{C}=[c_{ui}]_{M\times N}$ and computing policy $\textbf{D}=[d_{ui}]_{M\times N}$ by applying Algorithm 1.
        \State With given $\textbf{C}=[c_{ui}]_{M\times N}$ and $\textbf{D}=[d_{ui}]_{M\times N}$, find the optimal bandwidth allocation policy $\textbf{a}_u=[a_{u}]_{M\times 1}$ and the value of $Cost^{\text{max}}$.
        \EndWhile
	\end{algorithmic}
\end{algorithm}

\section{Simulation results}

In this section, simulation results are presented and discussed.
Without specific highlights, the simulation parameters are set as Table \ref{table2}.
The number of the users and contents are set to $M=5$ and $N=10$, respectively.
The gain, noise and bandwidth of the channel are set to $N_0=-174~\text{dBm/Hz}$, $h=3~\text{dB}$ and $B=20~\text{MHz}$, respectively.
At the same time, the ratio of the allocated power and bandwidth to each user from edge-cloud server is set to $pw=0.1~\text{Watt/MHz}$.
The power consumption of each user under idle state, transmission state and local executing state follow the uniform distribution between 0.001 - 0.009 Watt, 0.01 - 0.09 Watt and 0.1 - 0.5 Watt, respectively.
The weights of delay and energy consumption are set to $\lambda^e=0.2$ and $\lambda^t=0.8$, respectively.
The size of each 2D monocular video chunk is set to 3 Mbits while that of the corresponding 3D stereoscopic video chunk is set uniformly within [6, 8] Mbits in terms of contents.
For each content, the required processing density is set uniformly within [10, 20] cycles/bit.
The local cache capability of each user is set to 4 monocular video chunks.
The edge-cloud computing capability is set to 4 Gigacycles/s, and the local computing capability is set uniformly within [0.5, 1.5] Gigacycles/s.
In the following table, the ``Unchanged'' parameters are kept unchanged, while the ``Default'' parameters may change in simulation.

According to the Joint 3C optimization policy, all of the users should acquire the same cost value after optimization under the default value presented in Table \ref{table2}.
Thus, the subscript $u$ of $Cost_u$ can be ignored.



\begin{table}[h]\footnotesize
\centering
\begin{threeparttable}
\setlength{\abovecaptionskip}{0pt}
\caption{Simulation Parameter.}
\captionsetup{justification = centerlast}
\label{table2}
\begin{tabular}{p{1.2cm}<{\centering} | p{1.2cm}<{\centering} | p{4.2cm}<{\centering}}
\hline
 & Parameter & Value\\
\hline
\multirow{15}{*}{Unchanged} & $\lvert\mathcal{I}\rvert$ & 10\\
\multirow{15}{*}{} & $N_0$ & -174 dBm/Hz \cite{7842160}\\
\multirow{15}{*}{} & $pw$ & 0.1 Watt/MHz\\
\multirow{15}{*}{} & $p_u^{id}$ & 0.001 - 0.009 Watt uniformly \cite{6846368}\\
\multirow{15}{*}{} & $p_u^{com}$ & 0.01 - 0.09 Watt uniformly \cite{6846368}\\
\multirow{15}{*}{} & $p_u^{loc}$ & 0.1 - 0.5 Watt uniformly \cite{6846368}\\
\multirow{15}{*}{} & $h$ & 3 dB\\
\multirow{15}{*}{} & $\lambda^e$ & 0.2 \\
\multirow{15}{*}{} & $\lambda^t$ & 0.8\\
\multirow{15}{*}{} & $D^{pl}$ & 3 Mbits \cite{8728029}\\
\multirow{15}{*}{} & $D_i^{sp}$ & 6 - 8 Mbits uniformly \cite{8728029}\\
\multirow{15}{*}{} & $\sigma_i$ & 10 - 20 cycles/bit uniformly \cite{8319985}\\
\multirow{15}{*}{} & $\textcolor{blue}{f_u^{loc}}$ & 0.5 - 1.5 Gigacycles/s uniformly \cite{7517217}\\
\multirow{15}{*}{} & $Cost^{\text{max}}$ & 0.1\\
\multirow{15}{*}{} & $Cost^{\text{opt}}$ & -0.1\\
\hline

\multirow{4}{*}{Default} & $\lvert\mathcal{U}\rvert$ & 5 \\
\multirow{4}{*}{} & $C_u$ & 4\\
\multirow{4}{*}{} & $B$ & 30 MHz\\
\multirow{4}{*}{} & $f^c$ & 2 Gigacycles/s \cite{7914660}\\
\hline
\end{tabular}
\end{threeparttable}
\end{table}

\subsection{Analysis on request probability matrix}

On the basis of the Bert model identified in the previous section, the Amazon Review Data (2018) dataset \cite{ni2019justifying} will be used to derive the user request probability matrix for the subsequent simulation.
Most of the existing work fits the user request probability in a coarse-grained method via a uniform distribution or a Zipf distribution based on the popularity of the content, which neglects the variability of interests among different users.
The proposed policy for obtaining the user request probability  matrix is a fine-grained method that can effectively take into account the differences in interest among users, and can be closer to the subjective interests of users than the coarse-grained method.

\subsection{Convergence of Algorithm 3}

\begin{figure}[h!]
\centering
\includegraphics[width=1.0\linewidth]{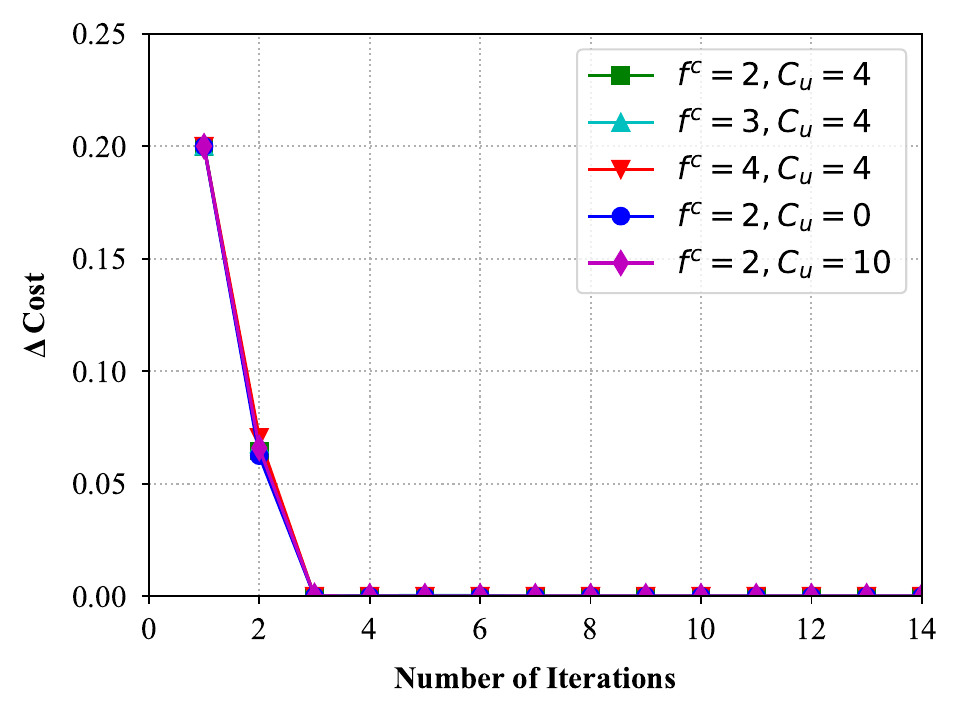}
\caption{Convergence of the main loop of Algorithm 3.}
\label{fig2}
\end{figure}

The convergence of $\text{Algorithm~3}$ is presented in Fig. \ref{fig2}.
For the sake of analysis and without loss of generality, denote $Cost^{\text{max}}=0.1$, $Cost^{\text{max}}=-0.1$ and $\Delta Cost = \lvert Cost^{\text{opt}} - Cost^{\text{max}} \rvert$, respectively, where $\Delta Cost$ donates the difference between the absolute values of $Cost^{\text{max}}$ and $Cost^{\text{opt}}$.
According to $\text{Algorithm~3}$, the while loop can be jumped out if and only if $\Delta Cost = 0$.
As can be seen in Fig. \ref{fig2}, the values of $\Delta Cost$ for all five cases listed reach 0 after almost three iterations.
In the accurate numerical results of the simulation, the number of iterations does not exceed 7.
Thus, the convergence of $\text{Algorithm~3}$ is verified.

\subsection{Analysis on the cache scheme}

\begin{figure}[h!]
\centering
\includegraphics[width=1.0\linewidth]{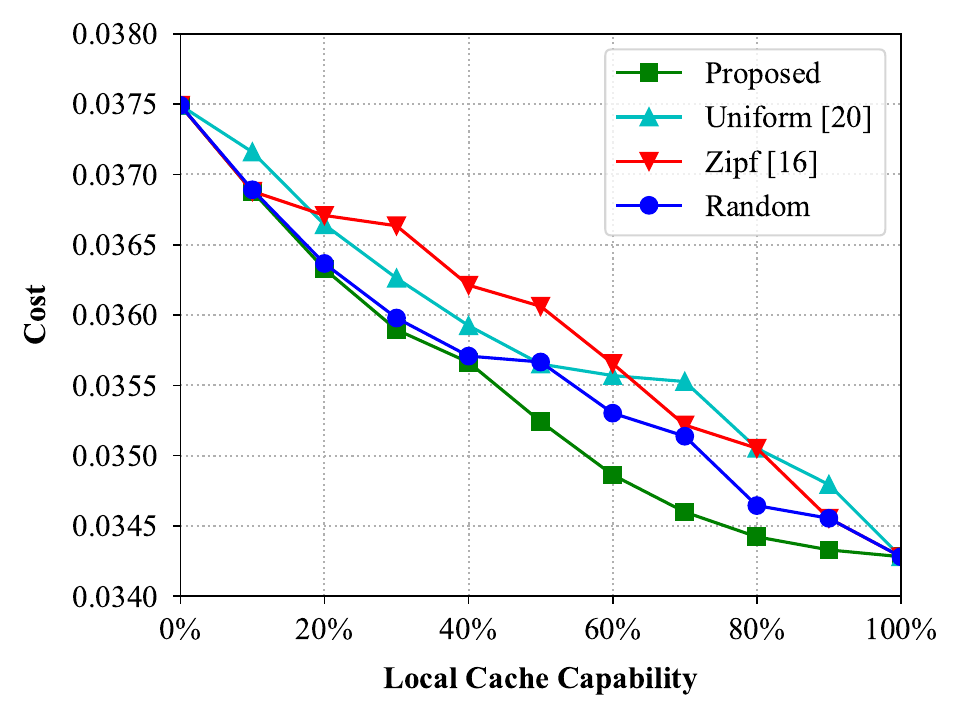}
\caption{Impacts of different caching schemes on VR service cost.}
\label{fig3}
\end{figure}

The impact of different caching schemes on VR service cost is depicted in Fig. \ref{fig3}.
In order to demonstrate the superiority of the proposed caching scheme, the following three caching schemes are taken into consideration for performance comparison.
The first caching scheme from work \cite{9013251} assumes that all contents share the same request probability and follow the uniform distribution.
The second caching scheme from work \cite{8728029} assumes that those contents may follow the Zipf distribution, i.e. the more popular content is, the higher request probability will be requested.
The third scheme assumes that the distribution of request probabilities for all contents is random.

\begin{figure*}[t!]
\centering
\subfigure[]
{
\begin{minipage}[b]{0.45\textwidth}
\centering
\includegraphics[width=1\textwidth]{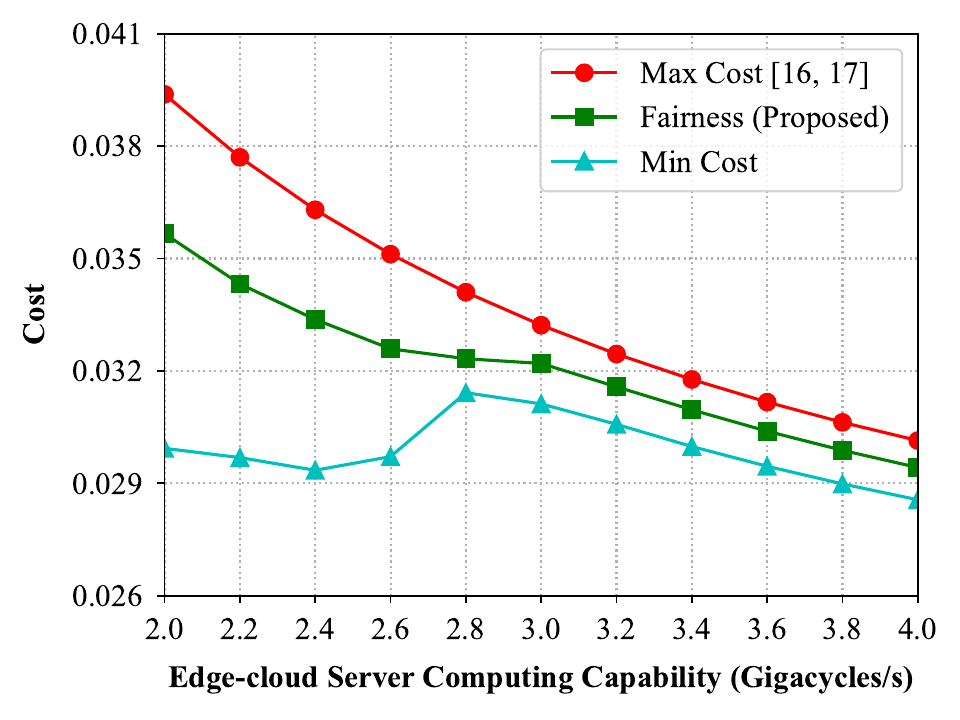}
\end{minipage}
\label{fig4_a}
}
\subfigure[]
{
\begin{minipage}[b]{0.45\textwidth}
\centering
\includegraphics[width=1\textwidth]{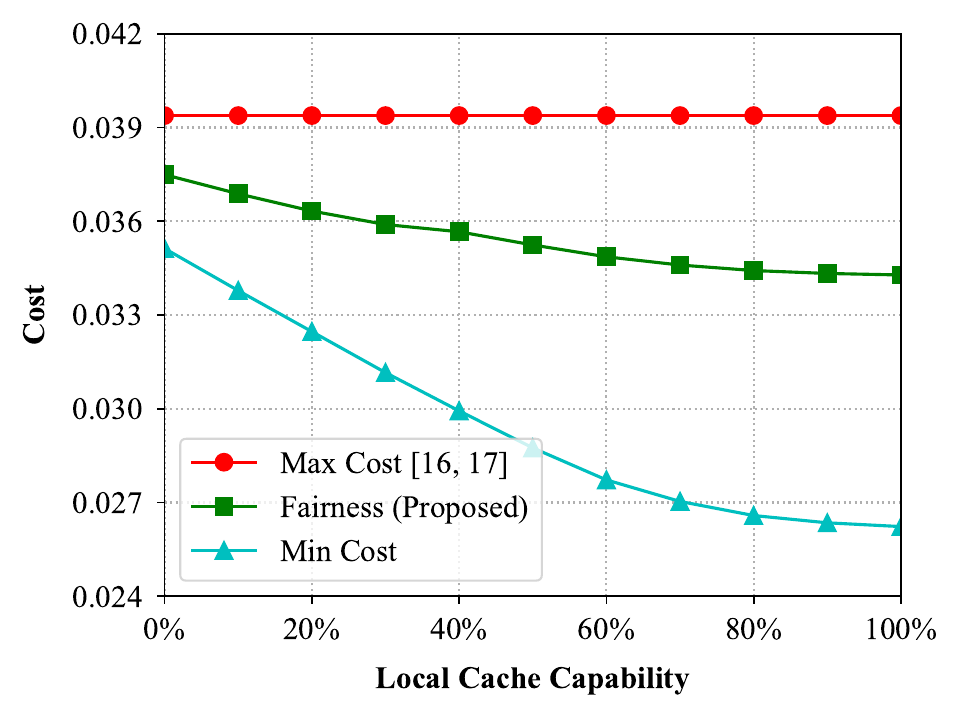}
\end{minipage}
\label{fig4_b}
}
\\
\subfigure[]
{
\begin{minipage}[b]{0.45\textwidth}
\centering
\includegraphics[width=1\textwidth]{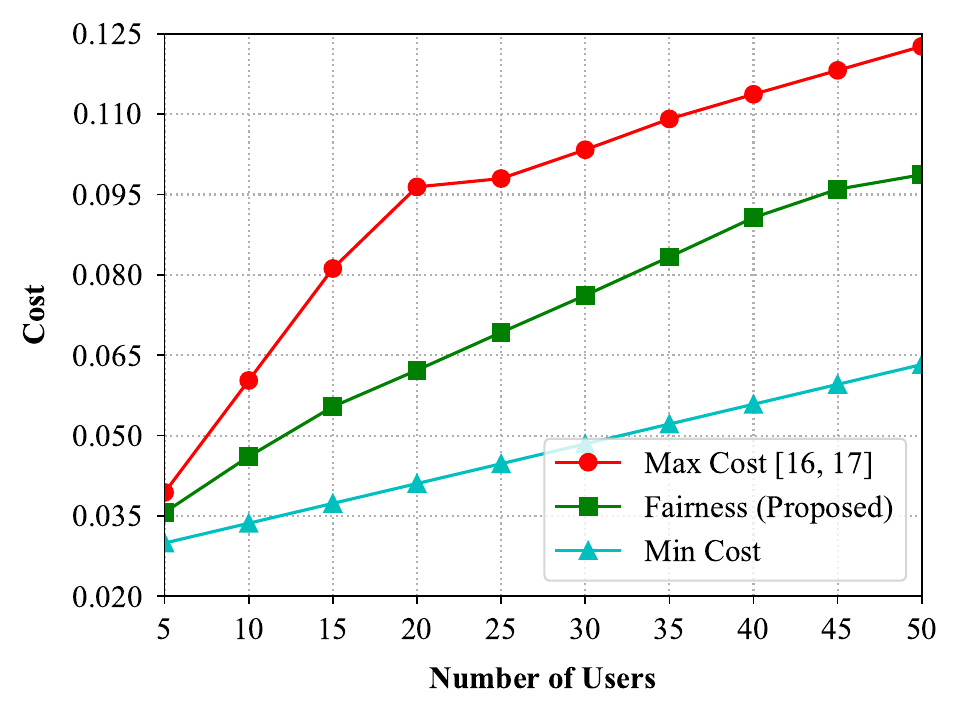}
\label{fig4_c}
\end{minipage}
}
\subfigure[]
{
\begin{minipage}[b]{0.45\textwidth}
\centering
\includegraphics[width=1\textwidth]{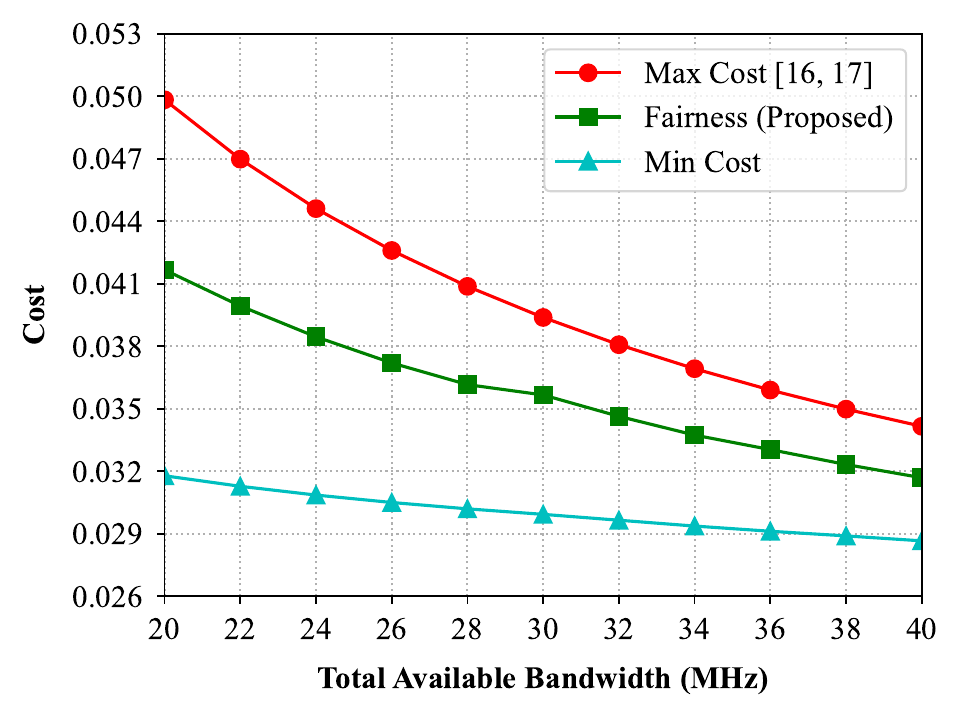}
\label{fig4_d}
\end{minipage}
}
\caption{Impacts of different parameters on user fairness performance. (a) Edge-cloud server computing capability. (b) Local cache capacity. (c) Number of users. (d) Total available bandwidth.}
\label{fig4}
\end{figure*}

Intuitively, as the local cache capability increase, local computing increasingly play an important role during the VR service, so that the VR service cost is gradually decreasing.
Among these four schemes, ours performs best since the others are just simple assumptions which are not based on the users' actual feelings.
When there are no local cache capacity in $UE$, none of those contents can be cached on the local side.
Therefore, no matter which caching scheme is chose, the final optimization results on VR service cost are the same.
On the other side, when $UE$ can cache all of the contents at the local device, which caching scheme to choose is no longer important, due to the fact that $UE$ can cache all of them locally.

\subsection{Analysis on user fairness}

The performance on user fairness under different parameters are described in Fig. \ref{fig4}.
The caching and computing policy should be firstly determined under the assumption that all users share the bandwidth resources equally throughout the whole optimization process.
In this case, the value of user's cost vary across all users though the optimal caching and computing policy are obtained.
Thus, the maximum and minimum cost value among all users are recorded as ``Max Cost'' and ``Min Cost'', respectively.

Due to multiple users in the small cell, cost value should be considered and optimized at the overall level, rather than individuals.
Hence, the performance of user fairness are taken into consideration in the proposed joint 3C optimization policy.
Similarly, the cost value of user optimized by proposed joint 3C optimization policy is recorded as ``Fairness''.

Since the enhancement of edge-cloud server computing capability and local cache capacity can improve the edge computing and the local computing ability, respectively, the optimal cost value declines continuously in Fig. \ref{fig4_a} and Fig. \ref{fig4_b}.
Compared to the value of ``Max Cost'' and ``Min Cost'', the proposed joint 3C policy has more performance improvement when the computing capability of the edge-cloud server is smaller in Fig. \ref{fig4_a} or the local cache capability of each user is bigger in Fig. \ref{fig4_b}.
This is due to the fact that when the computing capability at edge-cloud server is weaker or the local cache capability is stronger, more computations will be done locally.
Hence, the performances of VR service vary widely among different local devices and the proposed joint 3C policy can improve more VR service performance since their different local computing capabilities.

In addition, as the number of users grows, fewer bandwidth resources can be allocated to each user, leading to the increment of the cost value in Fig. \ref{fig4_c}.
Similarly, with the total available bandwidth increasing in Fig. \ref{fig4_d}, more bandwidth resources can be allocated, thus the cost value is decreasing.
This is owing to the fact that more available communication resources will increase the data transmission speed from the edge-server to the local side, which in turn will reduce the amount of computation performed locally, and the variability of performance among local devices will be gradually reduced affected by the heterogeneity of those local devices.

\begin{figure*}[t!]
\centering
\subfigure[]
{
\begin{minipage}[b]{0.45\textwidth}
\centering
\includegraphics[width=1\textwidth]{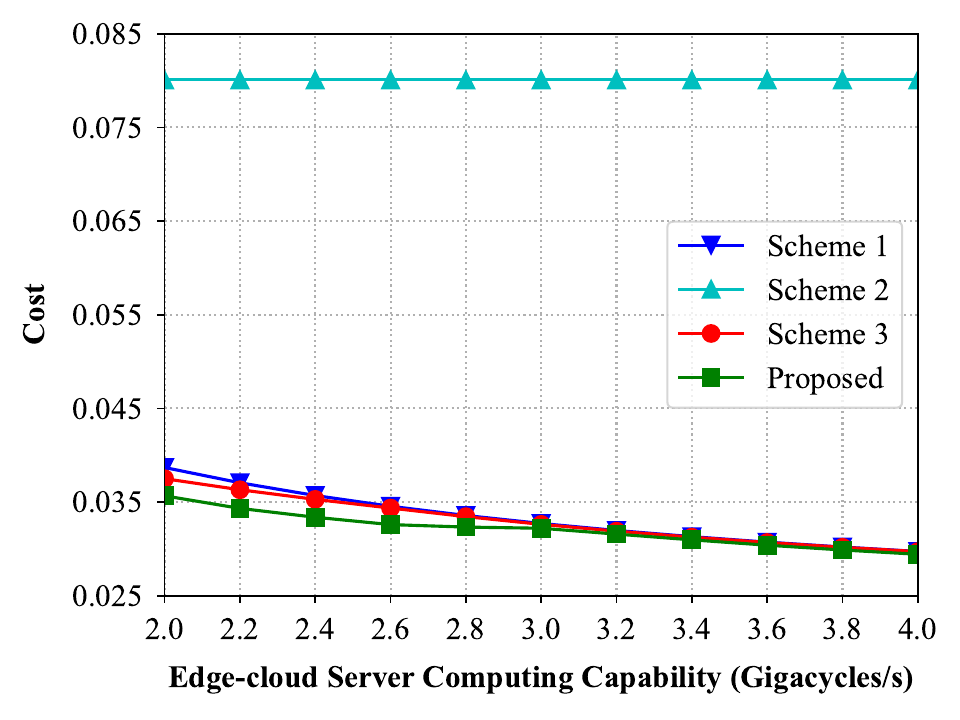}
\end{minipage}
\label{fig5_a}
}
\subfigure[]
{
\begin{minipage}[b]{0.45\textwidth}
\centering
\includegraphics[width=1\textwidth]{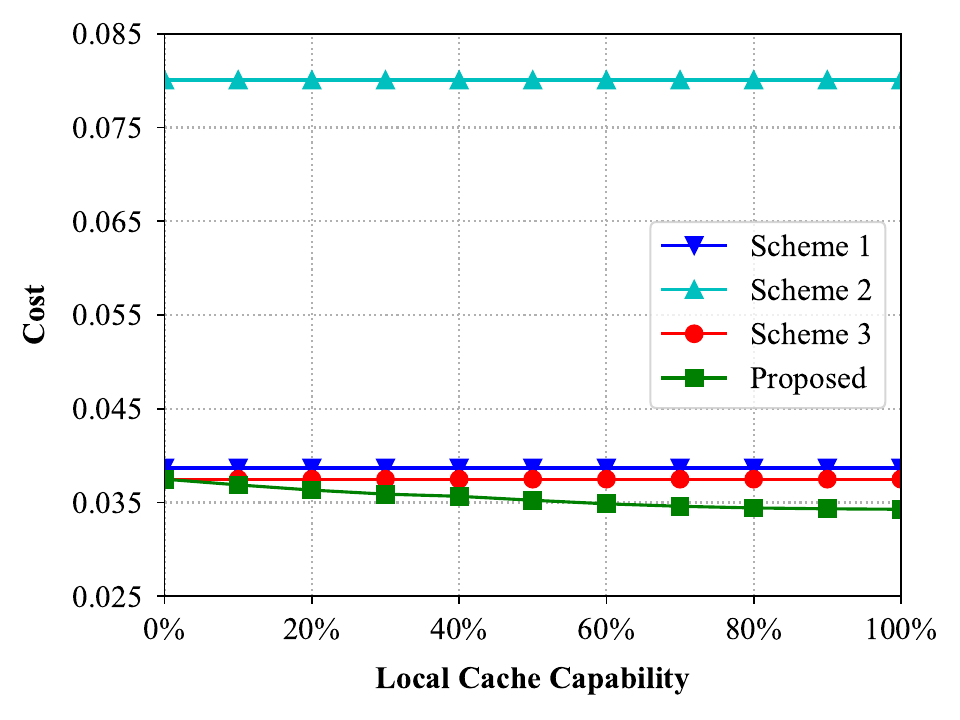}
\end{minipage}
\label{fig5_b}
}
\\
\subfigure[]
{
\begin{minipage}[b]{0.45\textwidth}
\centering
\includegraphics[width=1\textwidth]{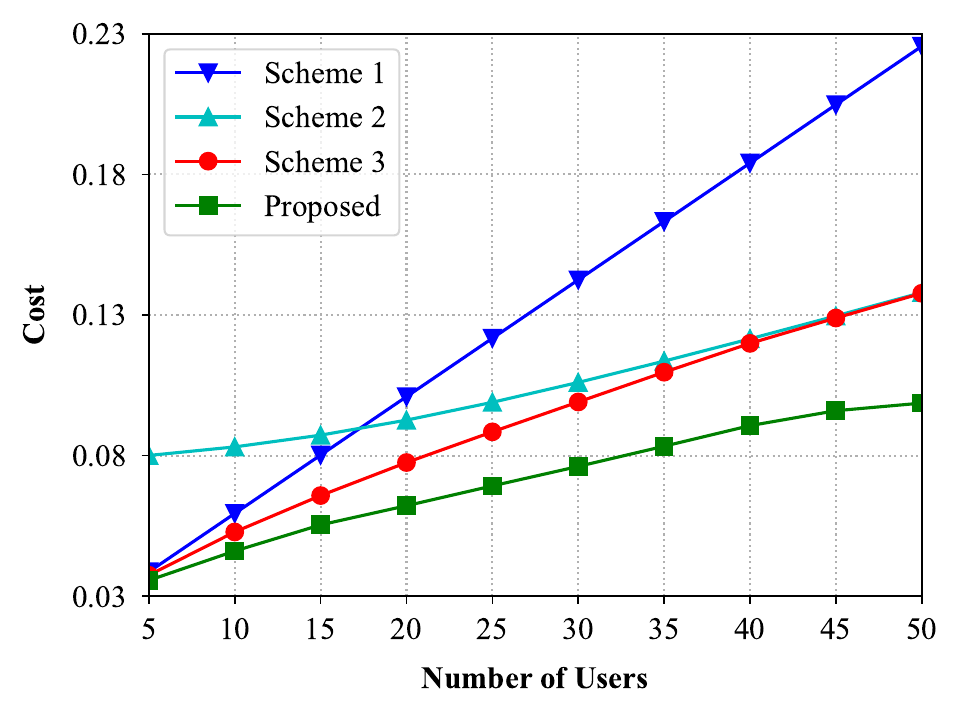}
\label{fig5_c}
\end{minipage}
}
\subfigure[]
{
\begin{minipage}[b]{0.45\textwidth}
\centering
\includegraphics[width=1\textwidth]{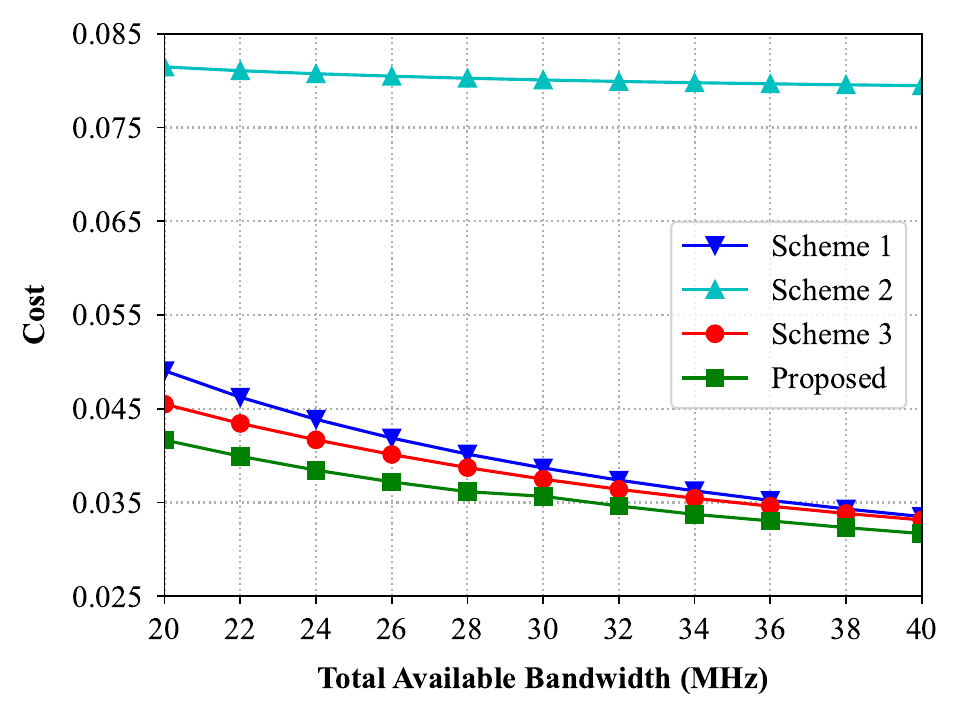}
\label{fig5_d}
\end{minipage}
}
\caption{Impacts of different schemes on VR service performance. (a) Edge-cloud server computing capability. (b) Local cache capacity. (c) Number of users. (d) Total available bandwidth.}
\label{fig5}
\end{figure*}

Overall, the proposed joint 3C optimization policy effectively improves the performance of the relatively worst-performing user in the small cell.
This prevents the extreme case from happening where good user performs better and poor user performs worse.

\subsection{Analysis on joint 3C optimization scheme}

The performance of the proposed joint 3C optimization policy is illustrated as shown in Fig. \ref{fig5}.
In order to validate the effectiveness of the proposed joint caching, computing and communication scheme, the following three other  baseline schemes are introduced for further performance comparison:
\begin{itemize}
\item Scheme 1 (Greedy edge computing): Contents are totally processed on the edge-cloud server without locally computing. After computing, those contents are totally transmitted to the local device.

\item Scheme 2 (Greedy local computing without caching): All of the 2D monocular video chunks are transmitted to the local device first in order to be computed on the $UE$ subsequently.

\item Scheme 3 (Joint 3C policy without caching): Those contents are transmitted and processed by edge-cloud server and $UE$ cooperatively. Contents can be computed on the edge-cloud server and then transmitted to the local device or transmitted to the device first to be computed locally.
\end{itemize}

From Fig. \ref{fig5_a}, it is observed that the proposed scheme performs best among all of the schemes.
As the edge-cloud server computing capability growing up, the performance of Scheme 2 remains unchange due to that the performance of local computing is independent of any parameters which related to edge-cloud server.
Also, when the edge-cloud server computing capability is strong, the performances of that three remaining schemes are closer.
This will result in more computation occurring on the edge-cloud server, rather than the local side.
At this point, whether the content is cached or computed locally is no longer important.

As depicted in  Fig. \ref{fig5_b}, since Scheme 2 does not consider the local caching, the performance of Scheme 2 still keeps unchange as the local cache capability grows.
When there is no local cache capacity, the lines of Scheme 3 and proposed scheme intersect at a point.
As the local cache capability grows, Scheme 3 outperforms Scheme 1 and Scheme 2, this demonstrates the superiority of joint 3C optimization policy over edge computing and local computing alone.
That performance of proposed scheme always performs better than Scheme 3 illustrates the superiority of local caching in end-to-edge collaboration system.
Local caching enables more contents to be processed locally, effectively reducing the transmission delay for edge computing and saving the communication resources in the system.

\begin{figure}[h!]
\centering
\subfigure[]{
\begin{minipage}{.8\linewidth}
\centering
\includegraphics[width=1\textwidth]{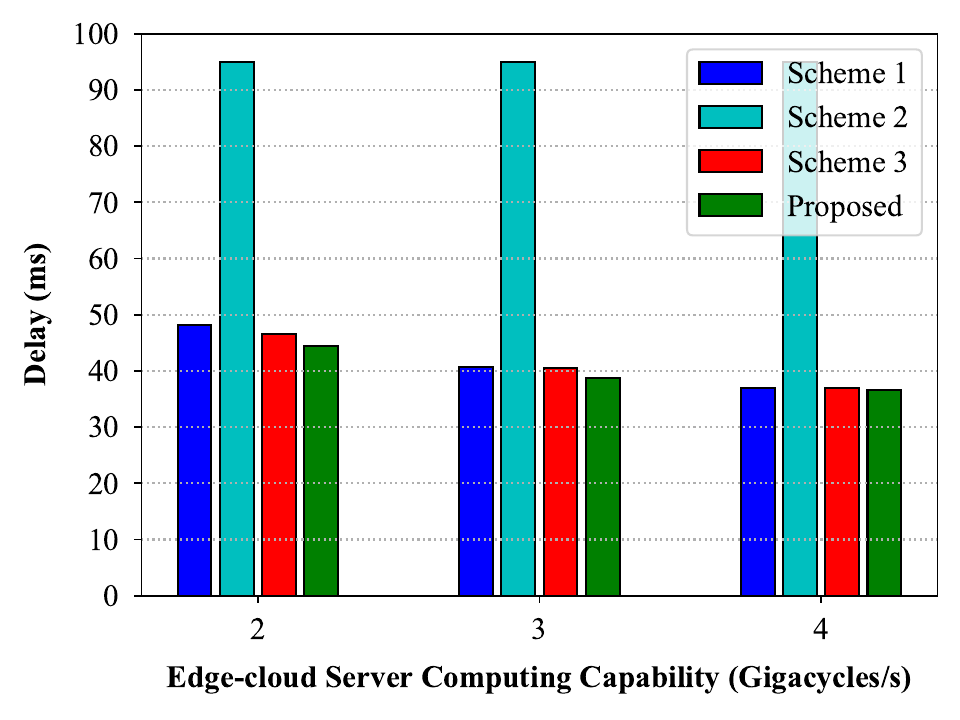}
\end{minipage}
\label{fig6_a}
}
\subfigure[]{
\begin{minipage}{.8\linewidth}
\centering
\includegraphics[width=1\textwidth]{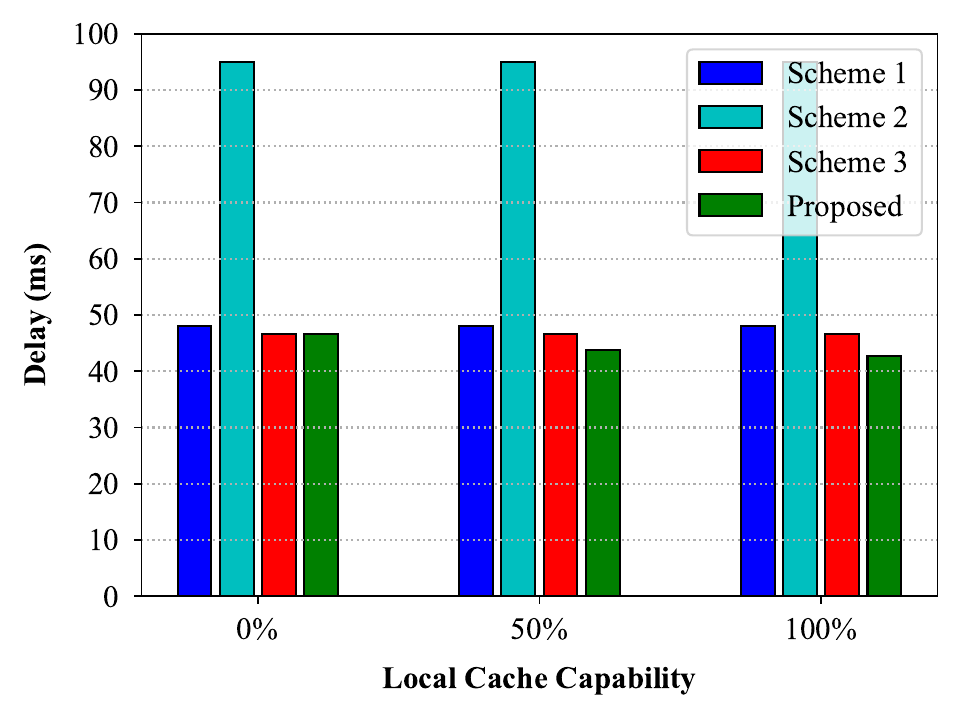}
\end{minipage}
\label{fig6_b}
}
\\
\subfigure[]{
\begin{minipage}{.8\linewidth}
\centering
\includegraphics[width=1\textwidth]{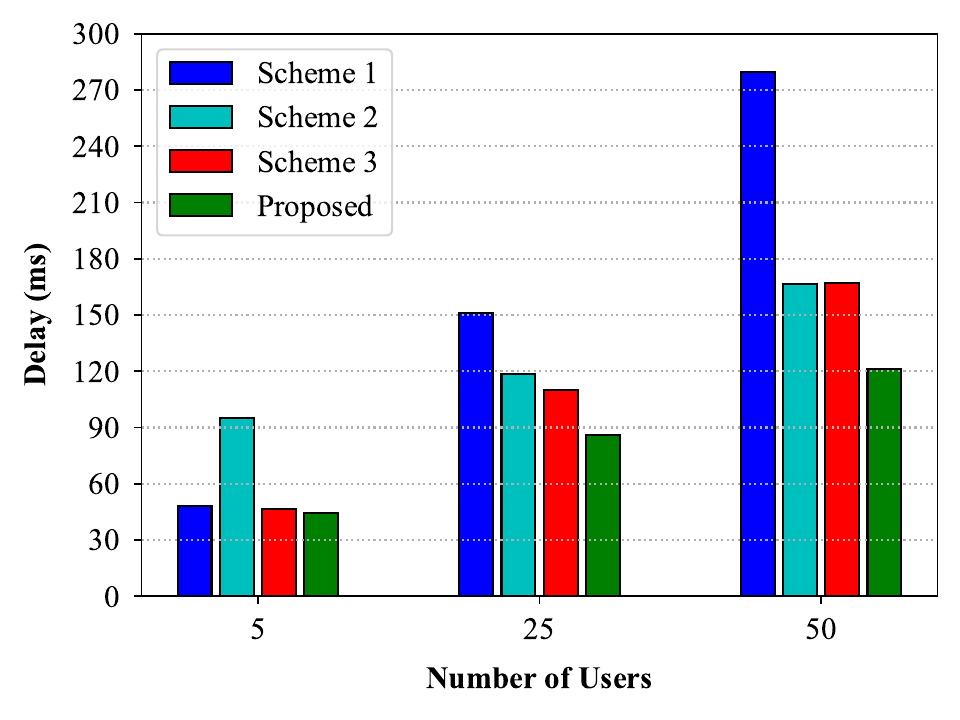}
\end{minipage}
\label{fig6_c}
}
\subfigure[]{
\begin{minipage}{.8\linewidth}
\centering
\includegraphics[width=1\textwidth]{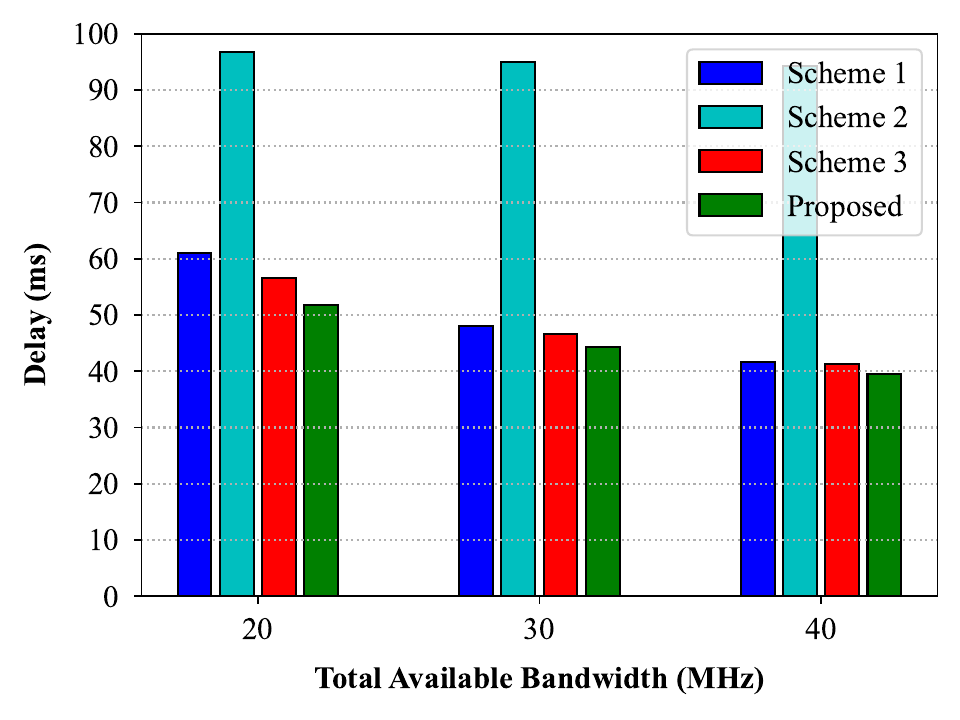}
\end{minipage}
\label{fig6_d}
}
\caption{Impacts of different schemes on VR service delay. (a) Edge-cloud server computing capability. (b) Local cache capacity. (c) Number of users. (d) Total available bandwidth.}
\label{fig6}
\end{figure}

Fig. \ref{fig5_c} depicts the variation of the number of users with optimal VR cost value.
Communication resources are relatively abundant when the number of users is small.
At this moment, edge computing performs better than local computing.
Thus, Scheme 1, Scheme 3 and proposed scheme have similar performance and all work better than Scheme 2 (local computing).
On the other hand, when the number of users is large, each user has a very limited share of the communication resources.
The cost value of processed contents transmitted to the local side from edge-cloud server will be relatively large when choosing edge computing.
Therefore, local computing may performs better than edge computing with a large number of users in the small cell, resulting an approximate intersection point of Scheme 2 and Scheme 3 when the number of users reaches 50 in Fig. \ref{fig5_c}.

An illustration of the variation of different schemes with the total available bandwidth is depicted in Fig. \ref{fig5_d}.
Since all schemes include the transfer of data from the edge-cloud server to the local side whether those data are raw 2D data to be processed or processed 3D data, any increase in total available bandwidth will improve the VR performance of all schemes.

\begin{figure}[h!]
\centering
\subfigure[]{
\begin{minipage}{.8\linewidth}
\centering
\includegraphics[width=1\textwidth]{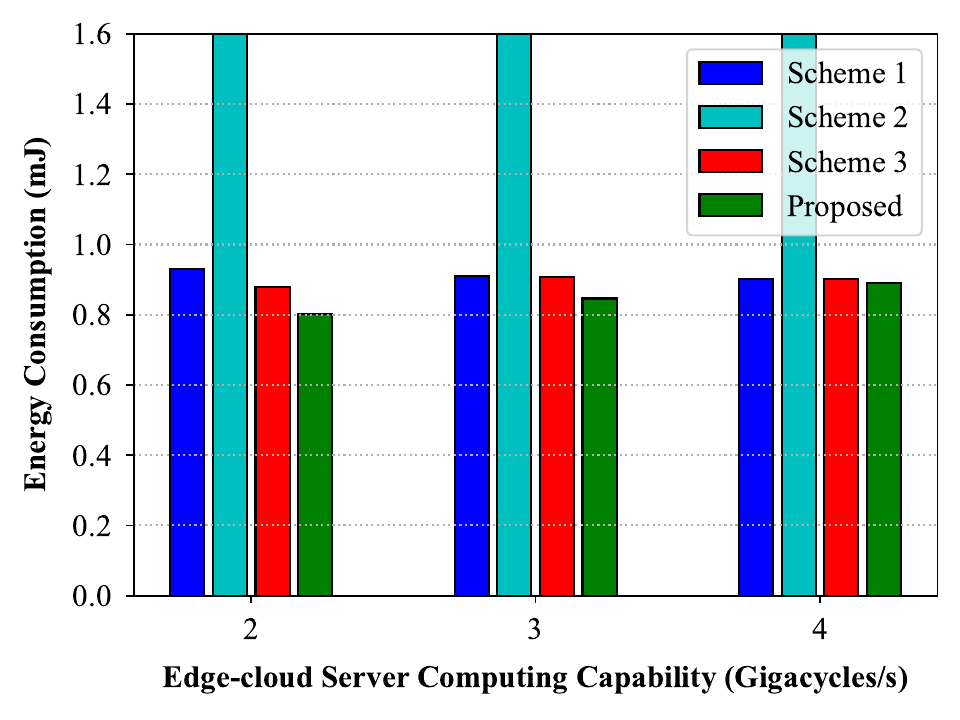}
\end{minipage}
\label{fig7_a}
}
\subfigure[]{
\begin{minipage}{.8\linewidth}
\centering
\includegraphics[width=1\textwidth]{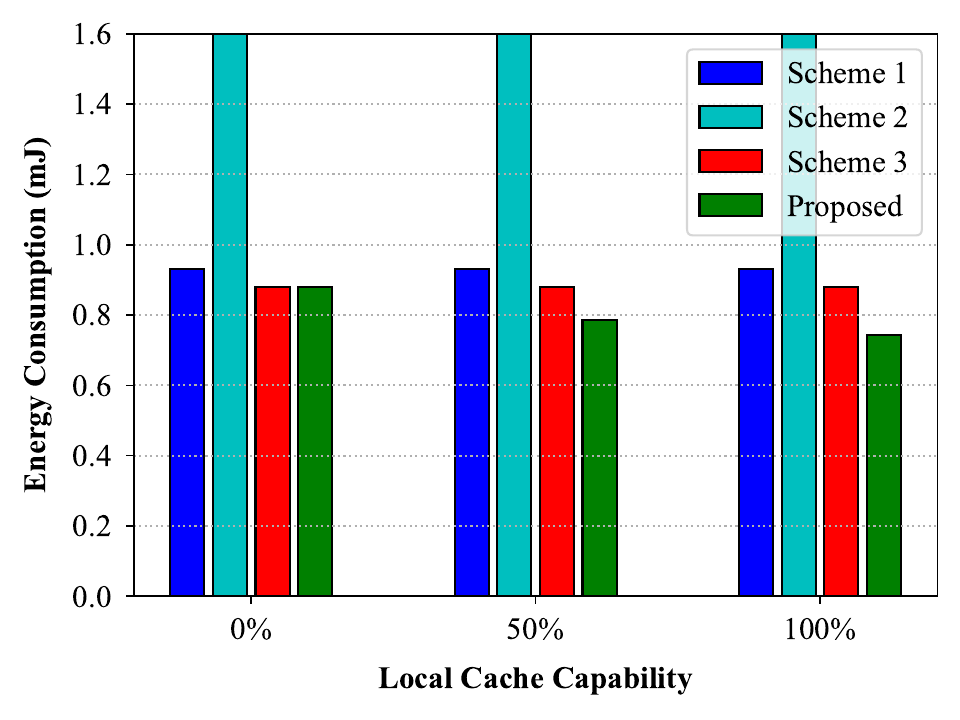}
\end{minipage}
\label{fig7_b}
}
\\
\subfigure[]{
\begin{minipage}{.8\linewidth}
\centering
\includegraphics[width=1\textwidth]{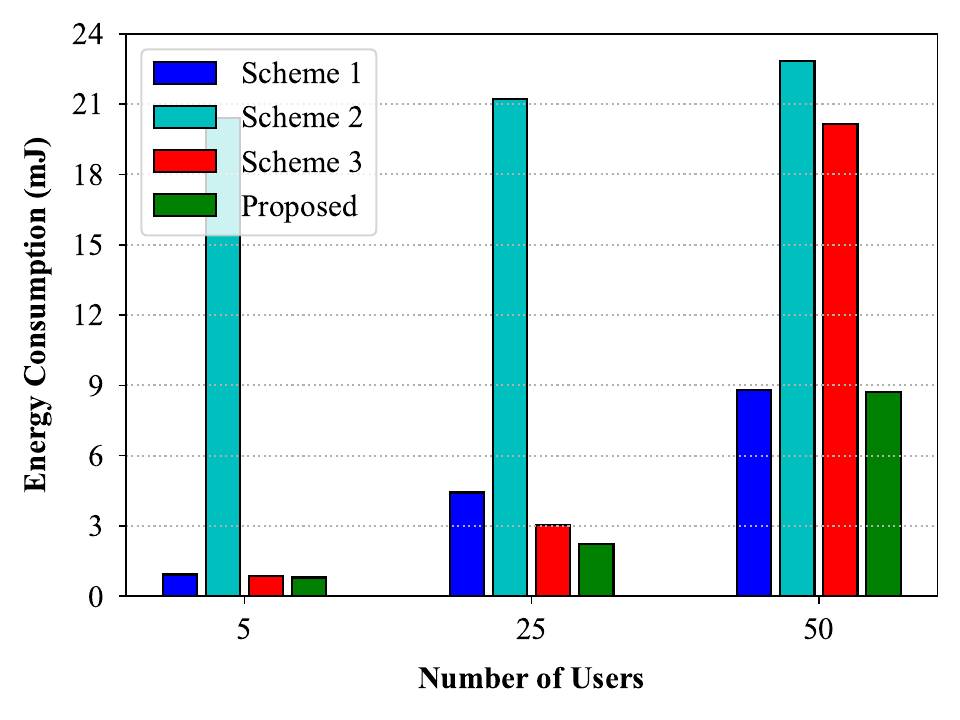}
\end{minipage}
\label{fig7_c}
}
\subfigure[]{
\begin{minipage}{.8\linewidth}
\centering
\includegraphics[width=1\textwidth]{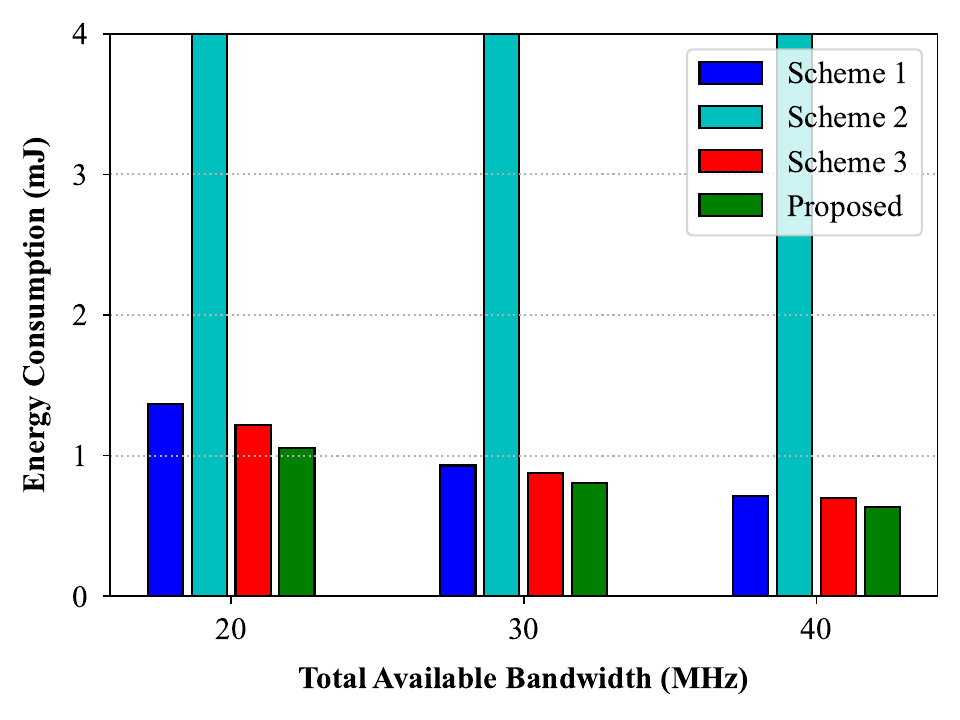}
\end{minipage}
\label{fig7_d}
}
\caption{Impacts of different schemes on local VR device energy consumption. (a) Edge-cloud server computing capability. (b) Local cache capacity. (c) Number of users. (d) Total available bandwidth.}
\label{fig7}
\end{figure}

Fig. \ref{fig6} and Fig. \ref{fig7} depict the variation of VR service delay and local VR device energy consumption in the network with respect to different influencing factors when optimizing the VR cost value, respectively.
Intuitively, the delay and energy consumption performance of our proposed scheme is the best among all those schemes.

Fig. \ref{fig6_a} illustrates the variation of VR service delay with edge-cloud server computing capability under different schemes.
As the computing power of the edge-cloud server increases, the performances of delay for all these schemes are gradually decreasing, except Scheme 2, which remains fixed.
This is due to the fact that Scheme 2 takes the form of local computation for VR services, any changes in the computational power of the edge-cloud server will have no impact on it.
Meanwhile, with the increasing computational power of edge servers, more computations will be placed on edge-cloud server to reduce delay, so the gap among the delay performance of Scheme 1, Scheme 3 and our proposed scheme will be reduced gradually.

Fig. \ref{fig6_b} describes the impact of the local cache capability on the delay of the VR service.
Since the first three schemes do not take into account the local cache, the delay performance of these three schemes does not vary with the local cache capacity.
When there is no local cache capacity, i.e., $C_u=0$, our proposed scheme will degrade into scheme 3, and both behaves the same performance in terms of VR service delay.

Fig. \ref{fig6_c} depicts the variation of VR service delay with the number of users.
It is intuitive that as the number of users increases in the network, fewer communication resources are allocated to each user, and thus the delay performance of each scheme increases.
Since the 3D data size for VR service is more than twice the size of 2D counterpart, the performance of Scheme 1 is affected much more than Scheme 2 as communication resources become scarce.
Thus, as the number of users increases in the network, Scheme 3 will gradually degenerate in to Scheme 2.

Similarly, Fig. \ref{fig6_d} portrays the impact of the total available bandwidth on the performance of VR service delay.
Due to the need for communication resources to transmit large amounts of 3D data, the delay performance of Scheme 1 is progressively improving as the total available bandwidth increases.
The VR service delay of Scheme 2 is composed of a computing delay and a communication delay for transmitting the 2D data.
Compared to the computing delay, the communication delay accounts for a relatively small amount, so the significant increase in communication resources can only slightly improve the overall VR service delay performance.

For the sake of convenience in describing the energy performance of remaining three schemes, the vertical coordinates of these graphs will be limited to a smaller range due to the excessive energy consumption of Scheme 2 in Fig. \ref{fig7}, where the value of energy consumption of Scheme 2 in Fig. \ref{fig7_a} and Fig. \ref{fig7_b} remains constant, and the value in Fig. \ref{fig7_d} is decreasing as the value of the horizontal coordinate increases.

In Fig. \ref{fig7_a}, increased edge-cloud server computing capability reduces computing delay, which leads to a reduction in the amount of time that local devices are in ``idle'' mode, which in turn reduces energy consumption in Scheme 1.
The anomalous slight increase in energy consumption in our proposed scheme is the result of the trade-off between delay and energy consumption during the optimization of VR cost value, where the weight of delay in the optimization objective of VR cost is higher than energy consumption.
Similar to Fig. \ref{fig6_b}, the change in the local cache capability in Fig. \ref{fig7_b} does not affect the computing and transmission process among Scheme 1, Scheme 2 and Scheme 3, and consequently will not have an impact on the energy consumption of these three schemes.
The energy consumption of our proposed scheme is decreasing as the local cache capability increases.

In Fig. \ref{fig7_c} and Fig. \ref{fig7_d}, as the number of users increases or the total available bandwidth decreases, the communication resources are decreasing and thus the increase in communication delay will result in higher local VR service energy consumption for all these four schemes.

\section{Conclusion}

This paper proposed a joint caching, computing and communication design to minimize the maximization VR service cost which can be characterized as a weighted sum of the overall VR delay and the energy consumption of local device under the guarantee of user fairness for an edge-terminal cooperative system.
The optimization problem is then decoupled into three independent subproblems to be solved separately.
With the proposed scheme, the performance of VR service is greatly improved compared to the other baseline schemes.
Furthermore, As the probability matrix for user request to content is derived by analyzing the collected realistic user comments, the proposed scheme is able to provide guideline on 3C resource allocation for VR service guarantee in practical networks.
Future work directions are listed as follows:

\begin{itemize}
\item Collaboration between users based on social ties in the same cell can be taken into consideration in future work.

\item The mobility of each user in the same cell can be taken into account in future work.

\item The scenario can be extended from one edge-cloud server to more than one edge-cloud server in one cell when interference management will be considered in future work.

\item The communication method for current work can be extended to semantic communication in future work, thus further reducing the amount of data required for communication and computation.

\end{itemize}

\section*{Acknowledgements}
This work was supported in part by the Graduate Research Innovation Project of Chongqing under grant CYB23237, and in part by the Doctoral Candidate Innovative Talent Program of CQUPT under grant BYJS202201, and in part by the National Natural Science Foundation of China (62271096, U20A20157), and in part by the Natural Science Foundation of Chongqing, China (cstc2020jcyjzdxmX0024), and in part by the University Innovation Research Group of Chongqing (CXQT20017), and in part by the Youth Innovation Group Support Program of ICE Discipline of CQUPT (SCIE-QN-2022-04).

%


\bibliographystyle{elsarticle-num}
\balance
\bibliography{egbib}

\begin{thebibliography}{10}
\expandafter\ifx\csname url\endcsname\relax
  \def\url#1{\texttt{#1}}\fi
\expandafter\ifx\csname urlprefix\endcsname\relax\def\urlprefix{URL }\fi
\expandafter\ifx\csname href\endcsname\relax
  \def\href#1#2{#2} \def\path#1{#1}\fi

\bibitem{ericsson}
Ericsson.com, Ericsson mobility report.
  \url{https://www.ericsson.com/en/reports-and-papers/mobility-report/reports/november-2022/},
  2022 (accessed 1 {D}ecember 2022).

\bibitem{meta}
Meta.com, Meta quest 3 is here. \url{https://www.meta.com/quest/quest-3/}, 2023
  (accessed 8 {O}ctober 2023).

\bibitem{playstation}
PlayStation.com, Immerse yourself in epic worlds that go beyond reality.
  \url{https://www.playstation.com/en-us/ps-vr2/}, 2023 (accessed 8 {O}ctober
  2023).

\bibitem{valve}
Store.steampowered.com, Valve index {VR} kit on steam.
  \url{https://store.steampowered.com/sub/354231/}, 2023 (accessed 8 {O}ctober
  2023).

\bibitem{vive}
Vive.com, Vive pro 2 - the best {VR} headset in the metaverse.
  \url{https://www.vive.com/us/product/vive-pro2/overview/}, 2023 (accessed 8
  {O}ctober 2023).

\bibitem{metapro}
Meta.com, Meta quest pro: Premium mixed reality.
  \url{https://www.meta.com/quest/quest-pro/}, 2023 (accessed 8 {O}ctober
  2023).

\bibitem{hp}
Hp.com, Hp reverb {G2} {VR} headset.
  \url{https://www.hp.com/us-en/vr/reverb-g2-vr-headset.html}, 2023 (accessed 8
  {O}ctober 2023).

\bibitem{huawei}
Huawei.com, Cloud {VR} network solution white paper.
  \url{https://www.huawei.com/minisite/pdf/ilab/cloud_vr_network_solution_white_paper_en.pdf},
  2018 (accessed 1 {J}anuary 2019).

\bibitem{rojas2023systematic}
M.~A. Rojas-S{\'a}nchez, P.~R. Palos-S{\'a}nchez, J.~A. Folgado-Fern{\'a}ndez,
  Systematic literature review and bibliometric analysis on virtual reality and
  education, Education and Information Technologies 28~(1) (2023) 155--192.

\bibitem{liu2022virtual}
Z.~Liu, L.~Ren, C.~Xiao, K.~Zhang, P.~Demian, Virtual reality aided therapy
  towards health 4.0: A two-decade bibliometric analysis, International journal
  of environmental research and public health 19~(3) (2022) 1525.

\bibitem{xi2021shopping}
N.~Xi, J.~Hamari, Shopping in virtual reality: A literature review and future
  agenda, Journal of Business Research 134 (2021) 37--58.

\bibitem{li2021study}
Y.~Li, H.~Song, R.~Guo, A study on the causal process of virtual reality
  tourism and its attributes in terms of their effects on subjective well-being
  during covid-19, International Journal of Environmental Research and Public
  Health 18~(3) (2021) 1019.

\bibitem{turkay2021virtual}
S.~Turkay, J.~Formosa, R.~Cuthbert, S.~Adinolf, R.~A. Brown, Virtual reality
  esports-understanding competitive players’ perceptions of location based
  {VR} esports, in: Proceedings of the 2021 CHI Conference on Human Factors in
  Computing Systems, 2021, pp. 1--15.

\bibitem{10175638}
M.~Mahmoud, S.~Rizou, A.~S. Panayides, N.~V. Kantartzis, G.~K. Karagiannidis,
  P.~I. Lazaridis, Z.~D. Zaharis, A survey on optimizing mobile delivery of
  360° videos: Edge caching and multicasting, IEEE Access 11 (2023)
  68925--68942.
\newblock \href {http://dx.doi.org/10.1109/ACCESS.2023.3292335}
  {\path{doi:10.1109/ACCESS.2023.3292335}}.

\bibitem{du2017computation}
J.~Du, L.~Zhao, J.~Feng, X.~Chu, Computation offloading and resource allocation
  in mixed fog/cloud computing systems with min-max fairness guarantee, IEEE
  Transactions on Communications 66~(4) (2017) 1594--1608.

\bibitem{8728029}
Y.~Sun, Z.~Chen, M.~Tao, H.~Liu, Communications, caching, and computing for
  mobile virtual reality: Modeling and tradeoff, IEEE Transactions on
  Communications 67~(11) (2019) 7573--7586.
\newblock \href {http://dx.doi.org/10.1109/TCOMM.2019.2920594}
  {\path{doi:10.1109/TCOMM.2019.2920594}}.

\bibitem{8713498}
T.~Dang, M.~Peng, Joint radio communication, caching, and computing design for
  mobile virtual reality delivery in fog radio access networks, IEEE Journal on
  Selected Areas in Communications 37~(7) (2019) 1594--1607.
\newblock \href {http://dx.doi.org/10.1109/JSAC.2019.2916486}
  {\path{doi:10.1109/JSAC.2019.2916486}}.

\bibitem{9693960}
H.~Xiao, C.~Xu, Z.~Feng, R.~Ding, S.~Yang, L.~Zhong, J.~Liang, G.-M. Muntean, A
  transcoding-enabled 360° {VR} video caching and delivery framework for
  edge-enhanced next-generation wireless networks, IEEE Journal on Selected
  Areas in Communications 40~(5) (2022) 1615--1631.
\newblock \href {http://dx.doi.org/10.1109/JSAC.2022.3145813}
  {\path{doi:10.1109/JSAC.2022.3145813}}.

\bibitem{8664595}
J.~Ren, G.~Yu, Y.~He, G.~Y. Li, Collaborative cloud and edge computing for
  latency minimization, IEEE Transactions on Vehicular Technology 68~(5) (2019)
  5031--5044.
\newblock \href {http://dx.doi.org/10.1109/TVT.2019.2904244}
  {\path{doi:10.1109/TVT.2019.2904244}}.

\bibitem{9013251}
T.~Dang, M.~Peng, Y.~Liu, C.~Liu, Joint bandwidth, caching, and computing
  resource allocation for mobile {VR} delivery in f-rans, in: 2019 IEEE Global
  Communications Conference (GLOBECOM), 2019, pp. 1--6.
\newblock \href {http://dx.doi.org/10.1109/GLOBECOM38437.2019.9013251}
  {\path{doi:10.1109/GLOBECOM38437.2019.9013251}}.

\bibitem{9893020}
M.-C. Lee, A.~F. Molisch, Optimal delay-outage analysis for noise-limited
  wireless networks with caching, computing, and communications, IEEE
  Transactions on Wireless Communications 22~(2) (2023) 1417--1431.
\newblock \href {http://dx.doi.org/10.1109/TWC.2022.3204738}
  {\path{doi:10.1109/TWC.2022.3204738}}.

\bibitem{10062546}
D.~Liu, Q.~Zheng, Y.~Shen, P.~Ding, M.~Cheriet, Multi-cell mec space partition
  and dynamic adjustment scheme for {VR} video transmission, in: 2022
  International Conference on Information Processing and Network Provisioning
  (ICIPNP), 2022, pp. 24--28.
\newblock \href {http://dx.doi.org/10.1109/ICIPNP57450.2022.00012}
  {\path{doi:10.1109/ICIPNP57450.2022.00012}}.

\bibitem{9839231}
M.-C. Lee, A.~F. Molisch, Asymptotic delay–outage analysis for noise-limited
  wireless networks with caching, computing, and communications, in: ICC 2022 -
  IEEE International Conference on Communications, 2022, pp. 4806--4811.
\newblock \href {http://dx.doi.org/10.1109/ICC45855.2022.9839231}
  {\path{doi:10.1109/ICC45855.2022.9839231}}.

\bibitem{9843968}
Z.~Yu, J.~Liu, C.~Wang, Q.~Yang, Bandit learning-based edge caching for
  360-degree video streaming with switching cost, IEEE Access 10 (2022)
  80714--80726.
\newblock \href {http://dx.doi.org/10.1109/ACCESS.2022.3194512}
  {\path{doi:10.1109/ACCESS.2022.3194512}}.

\bibitem{8332500}
J.~Park, P.~Popovski, O.~Simeone, Minimizing latency to support {VR} social
  interactions over wireless cellular systems via bandwidth allocation, IEEE
  Wireless Communications Letters 7~(5) (2018) 776--779.
\newblock \href {http://dx.doi.org/10.1109/LWC.2018.2823761}
  {\path{doi:10.1109/LWC.2018.2823761}}.

\bibitem{9667509}
Z.~Gu, H.~Lu, C.~Zou, Horizontal and vertical collaboration for vr delivery in
  mec-enabled small-cell networks, IEEE Communications Letters 26~(3) (2022)
  627--631.
\newblock \href {http://dx.doi.org/10.1109/LCOMM.2021.3140072}
  {\path{doi:10.1109/LCOMM.2021.3140072}}.

\bibitem{9268953}
Y.~Zhou, C.~Pan, P.~L. Yeoh, K.~Wang, M.~Elkashlan, B.~Vucetic, Y.~Li,
  Communication-and-computing latency minimization for uav-enabled virtual
  reality delivery systems, IEEE Transactions on Communications 69~(3) (2021)
  1723--1735.
\newblock \href {http://dx.doi.org/10.1109/TCOMM.2020.3040283}
  {\path{doi:10.1109/TCOMM.2020.3040283}}.

\bibitem{vaswani2017attention}
A.~Vaswani, N.~Shazeer, N.~Parmar, J.~Uszkoreit, L.~Jones, A.~N. Gomez,
  {\L}.~Kaiser, I.~Polosukhin, Attention is all you need, Advances in neural
  information processing systems 30.

\bibitem{maas2011learning}
A.~Maas, R.~E. Daly, P.~T. Pham, D.~Huang, A.~Y. Ng, C.~Potts, Learning word
  vectors for sentiment analysis, in: Proceedings of the 49th annual meeting of
  the association for computational linguistics: Human language technologies,
  2011, pp. 142--150.

\bibitem{kim2014convolutional}
Y.~Kim, Convolutional neural networks for sentence classification, arXiv
  preprint arXiv:1408.5882.

\bibitem{liu2016recurrent}
P.~Liu, X.~Qiu, X.~Huang, Recurrent neural network for text classification with
  multi-task learning, arXiv preprint arXiv:1605.05101.

\bibitem{joulin2016bag}
A.~Joulin, E.~Grave, P.~Bojanowski, T.~Mikolov, Bag of tricks for efficient
  text classification, arXiv preprint arXiv:1607.01759.

\bibitem{devlin2018bert}
J.~Devlin, M.-W. Chang, K.~Lee, K.~Toutanova, Bert: Pre-training of deep
  bidirectional transformers for language understanding, arXiv preprint
  arXiv:1810.04805.

\bibitem{7842160}
Y.~Mao, J.~Zhang, S.~H. Song, K.~B. Letaief, Power-delay tradeoff in multi-user
  mobile-edge computing systems, in: 2016 IEEE Global Communications Conference
  (GLOBECOM), 2016, pp. 1--6.
\newblock \href {http://dx.doi.org/10.1109/GLOCOM.2016.7842160}
  {\path{doi:10.1109/GLOCOM.2016.7842160}}.

\bibitem{6846368}
W.~Zhang, Y.~Wen, D.~O. Wu, Collaborative task execution in mobile cloud
  computing under a stochastic wireless channel, IEEE Transactions on Wireless
  Communications 14~(1) (2015) 81--93.
\newblock \href {http://dx.doi.org/10.1109/TWC.2014.2331051}
  {\path{doi:10.1109/TWC.2014.2331051}}.

\bibitem{8319985}
X.~Yang, Z.~Chen, K.~Li, Y.~Sun, N.~Liu, W.~Xie, Y.~Zhao,
  Communication-constrained mobile edge computing systems for wireless virtual
  reality: Scheduling and tradeoff, IEEE Access 6 (2018) 16665--16677.
\newblock \href {http://dx.doi.org/10.1109/ACCESS.2018.2817288}
  {\path{doi:10.1109/ACCESS.2018.2817288}}.

\bibitem{7517217}
X.~Lyu, H.~Tian, C.~Sengul, P.~Zhang, Multiuser joint task offloading and
  resource optimization in proximate clouds, IEEE Transactions on Vehicular
  Technology 66~(4) (2017) 3435--3447.
\newblock \href {http://dx.doi.org/10.1109/TVT.2016.2593486}
  {\path{doi:10.1109/TVT.2016.2593486}}.

\bibitem{7914660}
T.~Q. Dinh, J.~Tang, Q.~D. La, T.~Q.~S. Quek, Offloading in mobile edge
  computing: Task allocation and computational frequency scaling, IEEE
  Transactions on Communications 65~(8) (2017) 3571--3584.
\newblock \href {http://dx.doi.org/10.1109/TCOMM.2017.2699660}
  {\path{doi:10.1109/TCOMM.2017.2699660}}.

\bibitem{ni2019justifying}
J.~Ni, J.~Li, J.~McAuley, Justifying recommendations using distantly-labeled
  reviews and fine-grained aspects, in: Proceedings of the 2019 conference on
  empirical methods in natural language processing and the 9th international
  joint conference on natural language processing (EMNLP-IJCNLP), 2019, pp.
  188--197.

\end{thebibliography}
~~~\\
~~~\\







\end{document}